\documentstyle[preprint,tighten,aps,psfig]{revtex}

\def\vec#1{{\mbox{\boldmath$#1$}}}

\newcommand{\PP}{\mbox{$\vec{P}$}}
\newcommand{\p}{\mbox{$\vec{p}$}}
\newcommand{\q}{\mbox{$\vec{q}$}}
\newcommand{\pp}{\mbox{$\vec{p}'$}}

\newcommand{\k}{\mbox{$\vec{k}$}}

\newcommand{\r}{\mbox{$\vec{r}$}}
\newcommand{\s}{\mbox{$\vec{s}$}}

\newcommand{\si}{\mbox{$\vec{\sigma}$}}
\newcommand{\vgamma}{\mbox{$\vec{\gamma}$}}
\newcommand{\vxi}{\mbox{$\vec{\xi}$}}

\begin{document}

\preprint{BNL-HET-98/44, TTP 99-03, hep-ph/9901394}
\draft

\date{January 1999}

\title{Positronium $S$ state spectrum: 
         analytic results at ${\cal O}(m \alpha ^6)$ }

\author{ Andrzej Czarnecki\thanks{
  e-mail:  czar@quark.phy.bnl.gov}}
  \address{Physics Department, Brookhaven National Laboratory,\\
            Upton, NY 11973}
\author{ Kirill Melnikov\thanks{
  e-mail:  melnikov@particle.physik.uni-karlsruhe.de}}
  \address{Institut f\"{u}r Theoretische Teilchenphysik,\\
            Universit\"{a}t Karlsruhe, D--76128 Karlsruhe, Germany}
\author{ Alexander Yelkhovsky\thanks{
  e-mail:  yelkhovsky@inp.nsk.su}}
  \address{ Budker Institute for Nuclear Physics,\\    
            Novosibirsk, 630090, Russia}
\maketitle

\begin{abstract}
We present an analytic calculation of the ${\cal O}(m\alpha^6)$ recoil
and radiative recoil corrections to energy levels of positronium $nS$
states and their hyperfine splitting.  A complete analytic formula
valid to ${\cal O}(m\alpha^6)$ is given for the spectrum of $S$
states.  Technical aspects of the calculation are discussed in detail.
Theoretical predictions are given for various energy intervals and
compared with experimental results.

\vspace{0.5cm}

\noindent
{\em PACS numbers: 36.10.Dr, 06.20.Jr, 12.20.Ds, 31.30.Jv}
\end{abstract}

\newpage
\section{Introduction}

Spectroscopy of positronium (Ps) provides a sensitive test of bound
state theory based on the Quantum Electrodynamics (QED).  Because of
the small mass of electron and positron, the effects of strong and
weak interactions are negligible compared with the accuracy of present
experiments.  For this reason positronium represents a unique system
which can, in principle, be described with very high precision by
means of the QED only.  Tests of the QED predictions are made possible
by the very high experimental accuracy of positronium spectroscopy
\cite{Mills88}.

The gross spectrum of positronium is well described by the
Schr\"odinger equation with the Coulomb potential.  Energy levels
are 
\begin{equation} 
E(n) = -\frac {m\alpha^2}{4n^2},
\label{ELO}
\end{equation} 
where $n$ is the principal quantum number.
For the purpose of interpreting modern experiments
the precision of Eq.~(\ref{ELO}) is insufficient.  Corrections to the
energy levels can in part be described by the
Quantum Mechanics; however, for a complete description one has to
resort to the Quantum Field Theory (QFT). Unfortunately,  an
application of the QFT to the
bound states is difficult and special methods have to be devised 
\cite{Barbieri:1978mf,FGross,Lepage:1977gd,Caswell:1986ui}.

Various approaches to bound state calculations have been reviewed
e.g.~in \cite{SapYen}.  Here we focus on a method close to the
so--called Non-Relativistic Quantum Electrodynamics (NRQED)
\cite{Caswell:1986ui}, which is an effective field theory based on
the QED, for small energies and momenta.  Eq.~(\ref{ELO}) implies that
the characteristic velocity of the electron and positron in
positronium is of the order of the fine structure constant $\alpha \ll
1$.  It is appropriate to apply a non-relativistic approximation to
this system.

Recently much progress has been achieved in the framework of
non-relativistic effective theories, mainly by employing dimensional
regularization.  It has been shown \cite{Pineda:1997bj} that this
regularization procedure permits an exact separation of effects
arising at various characteristic energy scales.  Using that method,
which we will call dimensionally regularized NRQED (NRQED$_\epsilon$)
the complete energy spectrum of Ps has been reproduced to order
$m\alpha^5$ \cite{Pineda:1998kn}.  More recently, we have computed
$m\alpha^6$ corrections to the hyperfine splitting (HFS) of the Ps
ground state \cite{Czarnecki:1998zv}, confirming one of previously
obtained numerical results \cite{Ph}.  In the present paper we
generalize that result to all $S$ states, confirming \cite{PhPRL}, and
compute also their spin independent shift at ${\cal O}(m\alpha^6)$
(obtained numerically in \cite{PhPRL}). 

It is convenient to describe the energy of an $nS$ state of Ps by
dividing it up into the spin--averaged part and a part dependent on
the total Ps spin (hyperfine splitting):
\begin{equation}
E(J,n) = E_{\rm aver}(n) +\s_{+}\s_{-}E_{\rm hfs}(n), 
\label{param}
\end{equation}
where $J$ is the total spin value of the Ps
and $\s _{\pm}$ are the spins of the electron and  positron, respectively.
One finds: 
\begin{eqnarray}
J&=&1 \quad {\rm (triplet\; state):} \qquad 
                                 \s_{+}\s_{-}= +\frac{1}{4},
\nonumber \\
J&=&0 \quad {\rm (singlet\; state):} \qquad \s_{+}\s_{-}= -\frac{3}{4}.
\end{eqnarray}
Both the spin--averaged energy and the hyperfine splitting can be
represented by series in powers and logarithms of the fine structure
constant.  In the lowest order $E_{\rm aver}(n)=E(n)$ is given by
Eq.~(\ref{ELO}), and $E_{\rm hfs}(n) = {\cal O}(m\alpha^4)$.  

To order $m\alpha^5$ the results for $E_{\rm aver}$ and $E_{\rm hfs}$
were found in \cite{KK,Fulton,Gupta}.  Those corrections have several
sources: electron and positron charge radii and anomalous magnetic
moments, vacuum polarization, two-photon exchange, two-photon
annihilation and one-loop correction to the single-photon
annihilation.

Current accuracy of high precision experiments requires a complete
calculation of the ${\cal O}(m\alpha^6)$ corrections $\Delta E_{\rm
aver}$ and $\Delta E_{\rm hfs}$.  

The most precisely measured property of positronium is the 
ground state HFS, i.e.~the energy difference between
the two lowest states with total spin $1$ and $0$. 
Two best experimental values are
\begin{equation}
\Delta \nu \equiv  E(1^3S_1)-E(1^1S_0)
=203\;387.5(1.6)\;{\rm MHz},
\label{Mills}
\end{equation}
found in \cite{Mills1,Mills2} and
\begin{equation}
\Delta \nu =203\;389.10(0.74)\;{\rm MHz},
\label{Hughes}
\end{equation}
obtained  in \cite{Ritter}. 
Another quantity of the experimental interest is the 
energy difference of $2^3S_1$ and $1^3S_1$ states
\cite{Fee}:
\begin{equation}\label{Fee}
E(2^3S_1)-E(1^1S_1) = 1\;233\;607\;216.4(3.2)\; {\rm MHz}.
\end{equation}
The absolute accuracy of this measurement is clearly less impressive
than that of the hyperfine splitting. However, since $m\alpha^6 =
18.658\; {\rm MHz}$, a complete calculation of the energy levels at
this order is warranted.

At  order $m\alpha^6$ both  $\Delta E_{\rm
aver}$ and $\Delta E_{\rm hfs}$ can be written as
\begin{equation}
\Delta E  = \Delta E_{\rm rad}+\Delta E_{\rm annih}
+\Delta E_{\rm rad\;rec}+\Delta E_{\rm rec}.
\end{equation}
The logarithmic contributions at this order, ${\cal O}(m\alpha^6
\ln\alpha)$, present in the annihilation $\Delta E_{\rm annih}$ and
recoil $\Delta E_{\rm rec}$ corrections, were found first
\cite{Bodwin:1978ut,Caswell:1979vz}.  $\Delta E_{\rm rad}$ arises from
the radiative corrections to the Breit potential at ${\cal
O}(\alpha,\alpha^2)$ \cite{Brodsky:1966vn,BMR1}.  
The three, two, and one-photon annihilation
contributions giving $\Delta E_{\rm annih}$ were found in
\cite{Adkins:1988nd}, \cite{AAB}, and \cite{Adkins97,Hoang:1997ki},
respectively. The non-annihilation radiative recoil contributions
$\Delta E_{\rm rad\; rec}$ were calculated in \cite{STY,PhK}, while pure
recoil corrections $\Delta E_{\rm rec}$ were obtained  in
\cite{Ph,Caswell:1986ui,AS} for the HFS and in \cite{PhPRL} for
$E_{\rm aver}$.

In this paper we present an analytic calculation of the recoil and
radiative recoil corrections, $\Delta E_{\rm rad}$ and $\Delta E_{\rm
rad\; rec}$, to energy levels of arbitrary $nS$ positronium states.
The rest of this paper is organized as follows: in Section
\ref{sec:frame} we discuss our method in general terms.  Section
\ref{sec:hfs} is devoted to the calculation of the HFS. Many technical
details of this calculation are discussed there.  In Section
\ref{sec:lev} we present a calculation of the average energy $E_{\rm
aver}$.  It is very similar to HFS, except that some additional
operators contribute.  Also the ${\cal O}(m\alpha^6)$ radiative
recoil corrections are discussed.  Our results are summarized in
Section \ref{sec:sum}, where also an overview of the theoretical and
experimental situation is given and a complete analytic formula for
the $nS$ energy levels to order $m\alpha^6$ is presented.

\section{Framework of the calculation}
\label{sec:frame}
Before getting into details, let us describe the general framework of
our calculation of the ${\cal O}(m\alpha^6)$ corrections to energy
levels.

First, we calculate an on--shell scattering amplitude for
non-relativistic $(v \ll 1)$ particles to the needed order (the fact
that $v \sim \alpha$ in Ps serves as a counting rule for contributions
of various operators).  In addition to the leading, single Coulomb
exchange, this includes the relative ${\cal O}(v^2)$ Breit corrections
and also higher order ${\cal O}(v^4,\alpha v^3)$ terms.  This
non-relativistic amplitude is gauge invariant, and taken with a minus
sign provides the potential for non--relativistic particles.

Next, we use the ordinary quantum mechanical perturbation theory to
find the corrections due to that potential; as unperturbed states we
use the solutions of the Schr\"odinger equation with the Coulomb
potential.  We get the ${\cal O}(m\alpha^6)$ correction to energy
levels as the sum of the first order correction due to ${\cal
O}(v^4,\alpha v^3)$ perturbation and of the second order correction
due to the Breit Hamiltonian.  Previously, this scheme was used for
the calculation of the ${\cal O}(m\alpha^6 \ln\alpha)$ corrections to
the levels of $S$-states \cite{KMYlog} and of the ${\cal
O}(m\alpha^6)$ corrections to the levels of $P$-states
\cite{KMYp,KMYp2}.

In the present calculation the result of the non-relativistic
calculation is divergent.  This is because also the short-distance
(``hard'') corrections contribute.  They arise from virtual momenta
regions of the order of electron mass and  cannot be obtained from the
non--relativistic expansion.

Our calculation is performed in the spirit of NRQED.  We apply
dimensional regularization, which offers technical advantages over
more common techniques, based on the introduction of an intermediate
cut-off to separate the relativistic and non--relativistic momentum
regions.  Dimensional regularization makes the matching of the
low-scale effective theory and the complete QED extremely simple.  We
find that in the sum of the short and long--distance contributions the
singularities in the parameter\footnote{Throughout the paper, we use
the following notations: $D=4-2\epsilon$ and $d=3-2\epsilon$.} $\epsilon$
disappear and one arrives at a finite result.

The spinor algebra in dimensional regularization requires some
comments.  In order to obtain the energy shift due to an operator
${\cal O}_i$ one has to calculate the trace of the form $\mbox{Tr}
\left [ \Psi^\dagger {\cal O}_i \Psi \right ]$, where $\Psi$ is an
appropriate wave function.  The spinor parts of the relevant wave
functions are
$$
\Psi_P = \frac{1+\gamma_0}{2\sqrt{2}} \gamma_5,\qquad
\Psi_O = \frac{1+\gamma_0}{2\sqrt{2}} \vgamma  \vxi ,
$$
for para and orthopositronium states, respectively.  In the latter
case, $\vec \xi$ is the polarization vector (we average over its
directions). The traces are calculated in a standard way
in the $D$-dimensional space.  One encounters only even numbers of
$\gamma_5$ matrices, and we treat them as anticommuting.

Since the matrix elements involve the positronium
wave function, it is easiest to calculate for the ground state ($n=1$).
However, once the corrections to the ground state have been found,
there is a convenient way of finding them for excited states, with an
arbitrary value of the principal quantum number $n$.  Only the
non-relativistic contributions have a non-trivial dependence on $n$.
Their computation in dimensional regularization would be difficult.
However, this task is simplified using other regularizations.
Finally, we eliminate the cut-off dependence by requiring that for
$n=1$ the result matches the formula we found for $n=1$.  The freedom
of choosing the regularization scheme simplifies considerably this
part of the calculation.

\section{HFS of the positronium ground state}
\label{sec:hfs}
In this Section we present a calculation of the recoil corrections to
the Ps ground state, $\Delta_{\rm rec}E_{\rm hfs}$.  It is given as a
sum of soft (non-relativistic)
Eq.~(\ref{HFSnonrel}) and hard Eq.~(\ref{hard}) scales:
\begin{equation}
\Delta_{\rm rec} E_{\rm hfs} 
=  \Delta_{\rm nonrel} E_{\rm hfs} +\Delta_{\rm hard} E_{\rm hfs} 
=  m\alpha^6
               \left( - \frac{ 1 }{ 6 } \ln \alpha + \frac{331}{432}
               - \frac{ \ln 2 }{ 4 }
               - \frac{17\zeta(3)}{8\pi^2} +
               \frac{5}{12\pi^2} \right). 
\label{Ehfsrecground}
\end{equation}
Those two groups of contributions are computed, respectively, in
Sections \ref{sec:soft} and \ref{sec:hard}.  Further, in Section
\ref{sec:nhfs}, we find a
generalization of this result for radially excited states (arbitrary
$n$):
\begin{equation}
\Delta_{\rm rec} E_{\rm hfs}(n) = \frac {m \alpha^6}{n^3} \left [ 
-\frac {1}{6} \left ( \ln \frac {\alpha}{n}  +\Psi (n) + \gamma_E \right ) 
+ \frac {7}{12 n} - \frac {1}{2n^2 }
+ \frac{295}{432}
               - \frac{ \ln 2 }{ 4 }
               - \frac{17\zeta(3)}{8\pi^2} +
               \frac{5}{12\pi^2}
\right ],
\label{Ehfsrecn}
\end{equation}
where $\Psi(n)$ is the logarithmic derivative of the 
$\Gamma$-function and $\gamma_E\simeq 0.577216$ is the Euler constant.
The $n$ dependence of this result and its numerical 
value at $n=1$ are in agreement with \cite{Ph}. 

\subsection{Soft scale contributions}
\label{sec:soft}

We divide up the non-relativistic contributions to HFS into 6 parts:
tree level Coulomb and magnetic photon exchanges,  retardation,
one-loop operators, and second iteration of Breit Hamiltonian which
includes intermediate S and D wave states:
\begin{equation}
 \Delta_{\rm nonrel} E_{\rm hfs} 
= \Delta_{\rm C} E_{\rm hfs} +\Delta_{\rm M} E_{\rm hfs} 
+\Delta_{\rm ret} E_{\rm hfs} 
+\Delta_{\rm 1-loop} E_{\rm hfs} +\Delta_{\rm S} E_{\rm hfs}
+\Delta_{\rm D} E_{\rm hfs}.  
\end{equation}
These partial results, given in
  Eqs.~(\ref{DeltaC}, \ref{DeltaM}, \ref{Deltaret}, 
\ref{Delta1}, \ref{Swavehfs}, \ref{DeltaD2}) add up to
\cite{Czarnecki:1998zv} 
\begin{equation}
\Delta_{\rm nonrel} E_{\rm hfs} = \frac{ \pi\alpha^3 }{ 3 m^2 } \psi^2(0)
               \left( \frac{ 1 }{\epsilon } - 4 \ln(m\alpha) + \frac{331}{18} \right).
\label{HFSnonrel}
\end{equation}
In the remainder of this Section we discuss in detail how these
contributions are calculated.  

According to standard procedure \cite{GuptaQED} we identify the
on--shell scattering amplitude, taken with the minus sign, with the
matrix element of an interaction operator in the momentum
representation.  The soft scale contributions are calculated using the
time-independent ``old-fashioned'' perturbation theory and the Coulomb
gauge.  Since this technique is not very common, let us recall its
basic ingredients.  
Exchange of a Coulomb or magnetic photon is described, respectively, by
$-4\pi\alpha/\q^2$ or $-4\pi\alpha\alpha_i
\otimes \alpha_j (\delta_{ij}- q_i q_j/\q^2)/2|\q|$. In the latter
case, the denominator $2|\q|$ arises from the magnetic photon's phase
space element.  

An intermediate state
introduces the factor $(E-E_{\rm int}+i0)^{-1}$, where $E_{\rm int}$
is the energy of the intermediate state and $E$ is the total energy of
the process.

Dirac spinors are 
\begin{equation}
\label{u} 
u(\p) =
\sqrt{\frac{ 2\omega_{p} }{ \omega_{p}+m }} \Lambda_{+ }(\p) w,
\end{equation}
where $w$ denotes the four-spinor of a particle at rest;
projectors on the positive and negative electron energy states are
given by
$$
\Lambda_{\pm }(\p)=\frac{1}{2} \left( 1 \pm 
                \frac{ \vec{\alpha}\p+\beta m }{ \omega_{p} } \right),
\qquad \omega_{p} = \sqrt{\p^2+m^2}. 
$$ 
In an expression for the potential the projector
$\Lambda_{-}$ contributes an additional minus sign.  

We begin with the contributions of the tree level effective operators,
describing an exchange of the Coulomb or magnetic quanta. The tree
level operators, relevant for the ${\cal O}(m\alpha^6)$ calculation of
the HFS, arise as ${\cal O}(v^2)$ corrections to the Breit potential.

\subsubsection{Tree-level Coulomb photon exchange}
For the HFS we need
the spin-dependent part of the ${\cal O}(v^4)$ correction to the
Coulomb exchange (see Eq.~(\ref{eqa:coul})):
\begin{equation}\label{VC}
V_C(\p',\p) = -\frac{\pi\alpha}{ 16 m^4}
\frac{ [\si\p,\si\p'][\si'\p,\si'\p'] }{ \q^2 }.
\end{equation}
To calculate the spin part of the matrix element, we take the
trace with $d$-dimensional sigma--matrices and find (the factor $1/d$ 
in Eq.~(\ref{14}) arises from the average over  directions of the o-Ps
polarization vector)
\begin{eqnarray}
\frac{1}{ 2d } {\rm Tr} \left( \sigma_i [\si\p,\si\p'] \sigma_i
                            [\si'\p',\si'\p] \right)
&=&4\frac{ d-4 }{ d }
  \left[ \p'^2 \p^2 - (\p'\p)^2 \right],
\label{14}
\\
\frac{1}{ 2 } {\rm Tr} \left( [\si\p,\si\p'] [\si'\p',\si'\p] \right)
&=& 4 \left[ \p'^2 \p^2 - (\p'\p)^2 \right],
\end{eqnarray}
respectively for ortho and parapositronium.
Using 
\begin{equation}
\p'^2 \p^2 - (\p'\p)^2 = (\p'\p) \; \q^2 - (\p'\q)(\q\p),
\end{equation}
and noting that the average value of $\p'\p$ in an $S$--state
vanishes, we obtain the contribution of $V_C(\p,\p')$ to the ground
state HFS:
\begin{equation}
\Delta_C E_{\rm hfs} =  \left\langle V_C(\p',\p) \right\rangle \Big|_{S=0}^{S=1}
= - \frac{\pi\alpha}{dm^4}
\left\langle \frac{ (\p'\q)(\q\p) }{ \q^2 } \right\rangle.
\label{int}
\end{equation}
In Eq.~(\ref {int}) the  matrix element is to be calculated over the ground 
state wave function in $d$ dimensions:
$$
\left\langle f(\p,\p') \right\rangle \equiv 
\int {{\rm d}^d p \over (2\pi)^d}{{\rm d}^d p' \over (2\pi)^d}
\phi(p)\phi(p') f(\p,\p').
$$
Let us briefly explain how the integral in Eq.~(\ref {int}) is
calculated.  Although the integrand does not look complicated, the
difficulty is that the exact form of the wave function $\psi(r)$ in
$d$ dimensions is not known. Fortunately, it turns out to be
unnecessary.

There are two alternative ways to calculate this integral.  One is to
transform it to the coordinate space.  A divergence arises at $r=0$
and in the final result is proportional to the $d$-dimensional
$\psi(0)$; the remaining, finite part can be easily calculated in
$d=3$.\footnote{In general, also the derivative of the wave function
at the origin, ${\rm d}\psi(r)/{\rm d} r$ at $r=0$, can appear in the
divergent part of the integral.  However, the Schr\"odinger equation
relates it to $\psi(0)$.}

In the alternative approach we use the fact that the wave function
in Eq.~(\ref {int}) satisfies the $d$-dimensional Schr\"odinger
equation, which in the momentum space reads\footnote{This equation
corresponds to a summation of an infinite number of ladder diagrams in
the Coulomb gauge.  For consistency it is essential to use here
dimensional regularization  in the same way as in the
other loop integrations.}
\begin{equation}
\phi(p) = \frac {4\pi \alpha m}{\p^2-mE}
\int \frac {{\rm d}^dk}{(2\pi)^d} 
\frac {\phi(k)}{(\p-\k)^2}.
\end{equation}
Using this equation we rewrite the integral in Eq.~(\ref {int}) as
\begin{equation}
\left\langle \frac{ (\p'\q)(\q\p) }{ \q^2 } \right\rangle _{\p,\p'} =  
\left\langle \frac{(4\pi\alpha m)^2(\p'\q)(\q\p) }
{(\p^2-mE)(\p-\k)^2 \q^2 (\p'^2-mE)(\p'-\k')^2}
\right\rangle _{\k,\k'},
\label {int1}
\end{equation}
where the integration over $\p$, $\p'$, as well as $\k$, $\k'$, in the
last expression is 
understood.  The integral over $\p$ and $\p'$ receives a divergent
contribution only from the region where $\p$ and $\p'$ simultaneously
become infinite. Therefore, a single subtraction is sufficient to make
this integral finite.  It is convenient to subtract from (\ref{int1})
the following expression: 
\begin{equation}
\left\langle \frac{(4\pi \alpha m)^2  (\p'\q)(\q\p) }{(\p^2-mE)^2 \q^2 (\p'^2-mE)^2}
\right\rangle _{\k,\k'}.
\label {int2}
\end{equation}
After the subtraction is done, two nice features emerge.  In Eq.~(\ref
{int2}) the integration over $\k,\k'$ factorizes and leads to
$\psi^2(0)$ times a two-loop integral, which can be easily calculated
for arbitrary $d$.  On the other hand, the difference between the last
integral in Eq.~(\ref {int1}) and the integral in Eq.~(\ref {int2}) is
finite and can be calculated for $d=3$ using the explicit form of the
wave function,
$$
\phi(p) = \sqrt{\pi\alpha m\over 2}{2m^2\alpha^2\over (\p^2-mE)^2},
\qquad
E=-{m\alpha^2\over 4}.
$$
We note that the counterterm (\ref {int2}) is
constructed in such a way that the above mentioned difference vanishes
for the ground state. This can be easily seen by integrating over
$\k,\k'$ in Eq.~(\ref {int2}) and using the fact that the
$\p,\p'$-dependent terms in the denominator of Eq.~(\ref {int2})
coincide (up to a normalization factor) with the three-dimensional
ground state wave functions in the momentum representation.

Both methods described above lead to the same result.  For $d = 3 -
2\epsilon$ we obtain:\footnote{We neglect factors
$\Gamma^2(1+\epsilon)$ and $(4\pi\mu^2)^{2\epsilon}$ which do not
contribute to the final, finite result.}
\begin{equation}
\Delta_C E_{\rm hfs} = \frac{ \pi\alpha^3 }{ 24 m^2 } \psi^2(0)
                \left( \frac{ 1 }{\epsilon } - 4 \ln(m\alpha) - \frac{1}{3} \right),
\label{DeltaC}
\end{equation}
where $\psi(0)$ is the value of the $d$-dimensional ground state wave
function at the origin.

\subsubsection{Tree-level exchange of a magnetic photon}
We now consider the correction caused by the tree level exchange of a
magnetic photon, Fig.~\ref{fig:soft}(b).  We neglect the energy
dependence in the photon propagator; it will be restored in the
following Section, where we discuss retardation effects. The relevant
potential is obtained from Eq.~(\ref{eqa:magn}):
\begin{equation}
\label{VM}
V_M(\p',\p)
             = \frac{ \pi\alpha }{ 16m^4 } 
               \frac{[\si'\q,\sigma'^i]}{ \q^2 } \left\{ 
               \left[ \si\frac{ \p'+\p }{ 2 }, \sigma^i \right]
\left(\p'^2-\p^2\right)
               + [\si\q,\sigma^i]\left(\p^2+\p'^2\right) \right\}
               + (\si \leftrightarrow \si').
\end{equation}
Contribution of this interaction to the ground state HFS is 
\begin{equation}\label{EM}
\Delta_M E_{\rm hfs} = \left\langle V_M(\p',\p) \right\rangle \Big|_{S=0}^{S=1}
= -2 \frac{d-1}{d} \frac{ \pi\alpha }{ m^4 }
  \left\langle \p^2 + \p'^2 + \frac{ \left(\p^2-\p'^2\right)^2 }{ 2\q^2 } \right\rangle.
\end{equation}
In $d=3$ this matrix element is linearly divergent. 
To demonstrate how we treat linear divergences let us consider the
$\p^2$ term on the RHS of the above equation: 
\begin{equation}
\left\langle \p^2 \right\rangle = \psi(0) 
\int \frac{{\rm d}^d p}{(2\pi)^d} \p^2 \phi(p)
= m \psi(0)  
\int \frac{{\rm d}^d p}{(2\pi)^d} \left( E \phi(p) 
+  
\int \frac{{\rm d}^d k}{(2\pi)^d} \frac{ 4\pi\alpha }{ (\p-\k)^2 } 
\phi(k) \right).
\label {tadpole}
\end{equation}
Shifting the integration variable $\p \to \p+\k$ we find that the
$\p$-integral in the last term is scale-less.  In dimensional
regularization such integrals vanish.  The first term in Eq.~(\ref
{tadpole}) is finite in three dimensions. We obtain
\begin{equation}
\left\langle \p^2 \right\rangle = mE \psi^2(0).
\end{equation}
Applying a similar procedure to the last term in Eq.~(\ref{EM}) 
we find the contribution of $V_M(\p',\p)$ to the ground state HFS:
\begin{equation}
\Delta_M E_{\rm hfs} = \frac{ \pi\alpha }{ m^4 }
               \left[ m^2\alpha^2 \psi^2(0) 
               - 4\frac{d-1}{d} \left\langle \frac{ (\p'\q)(\q\p) }{ \q^2 } \right\rangle
               \right] 
            = \frac{ \pi\alpha^3 }{ 3 m^2 } \psi^2(0)
                \left( \frac{ 1 }{\epsilon } - 4 \ln(m\alpha) + \frac{5}{3} \right).
\label{DeltaM}
\end{equation}

\subsubsection{Retardation effects}
Let us now consider the retardation effects, which mean that the
magnetic photon emitted by the electron propagates for a finite amount
of time before being absorbed by the positron. During this time, the
electron and positron can interact by several Coulomb exchanges
(Fig.~\ref{fig:soft}(c,d,e)).  To calculate the influence of these
effects on the HFS, it is sufficient to take the spin-dependent parts
of the current $\vec{j}(\pp, \p) = u^+(\pp)\vec{\alpha} u(\p)$ 
in the leading nonrelativistic approximation:
\begin{equation}
\vec{j}(\pp, \p) \to \frac{ [\si\q,\si] }{ 4m }.
\end{equation}
The scattering operator describing the retardation effects
is nonlocal both in space and time:
\begin{equation}\label{Aret}
- A_{\rm ret} = - \alpha 
\int \frac{{\rm d}^d k}{(2\pi)^d} 
\exp \left( -i\k\r_p \right) \frac{ [\si'\k,\sigma'_i] }{ 4m }
\frac{ 4\pi }{ 2k }\frac{ \delta_{ij} - \frac{k_i k_j}{\k^2} }{ k+H-E } 
\frac{ [\si\k,\sigma_j] }{ 4m }
\exp \left( i\k\r_e \right) + {\rm H.c.}
\label{retar}
\end{equation}
Here we assume that the magnetic photon with the momentum $\k$ is emitted 
by the electron at a point $\r_e$
and  absorbed by the positron at a point $\r_p$. 
Between those
moments, the evolution of the system ``positronium + photon'' is governed by
the propagator $4\pi(\delta_{ij}-k_i k_j/k^2)/(2k)(k+H-E)^{-1}$, $H$ being
the Hamiltonian of the nonrelativistic positronium slowly moving due to recoil.
In the region of interest $k \gg E$ and one can expand the amplitude
(\ref{Aret}) over the powers of $(H-E)/k\sim \alpha$. The zeroth term
of this expansion is the spin--dependent part of the Breit
potential,
\begin{equation}\label{A0ret}
- A_{\rm ret}^{(0)}(\q) = - \frac{ \pi\alpha }{ 4m^2 }
\frac{ [\si'\q,\sigma'^i][\si\q,\sigma^i] }{ \q^2 }.
\end{equation}
We need the second order term:
\begin{equation}\label{Vretpos}
V_{\rm ret} = \alpha 
\int \frac{{\rm d}^d k}{(2\pi)^d} \frac{ 4\pi }{ 2\k^4 }
\frac{ [\si'\k,\sigma'^i] }{ 4m }[H,\exp \left( -i\k\r_p \right)] 
[H,\exp \left( i\k\r_e \right)]\frac{ [\si\k,\sigma^i] }{ 4m }
 + {\rm H.c.}
\end{equation}
Only kinetic part of the Hamiltonian, 
\begin{equation}
H_{\rm kin} = \frac{ \p_e^2 }{ 2m } + \frac{ \p_p^2 }{ 2m },
\end{equation}
has to be retained in the commutators.
We find
\begin{equation}
V_{\rm ret} = - \alpha 
\int \frac{{\rm d}^d k}{(2\pi)^d} \frac{ 4\pi }{ 2\k^4 }
\frac{ [\si'\k,\sigma'^i] }{ 8m^2 }(\k^2+2\k\p_p)\exp \left
( i\k(\r_e-\r_p) \right)  
(\k^2+2\k\p_e)\frac{ [\si\k,\sigma^i] }{ 8m^2 }
 + {\rm H.c.}
\end{equation}
Transforming back to the relative coordinate $\r=\r_e-\r_p$ and the relative 
momentum $\p=\p_e=-\p_p$, we get for
the ground state HFS:
\begin{equation}
\Delta_{\rm ret} E_{\rm hfs} = \frac{ \pi\alpha }{ m^4 }
                       \left( \frac{m^2\alpha^2}{3} \psi^2(0) 
                       - 4\frac{d-1}{d} \left\langle \frac{
 (\p'\q)(\q\p) }{ \q^2 } \right\rangle \right)
= 
\frac{ \pi\alpha^3 }{ 3 m^2 } \psi^2(0)
                \left( \frac{ 1 }{\epsilon } - 4 \ln(m\alpha) - \frac{1}{3} \right).
\label{Deltaret}
\end{equation}

\subsubsection{One-loop operators}
Now we turn to the operators generated by 
one-loop diagrams. For the HFS the only contribution comes from the
graph in Fig.~\ref{fig:soft}(f), which describes the mixed
Coulomb-magnetic exchange with a transition of one of the particles
to a negative energy state.  In other words, this corresponds to a
creation of an additional electron-positron pair by the electric or
magnetic field of the electron or positron.

Using Feynman rules for the time-independent perturbation
theory, given at the beginning of this Section, we derive
the corresponding potential:
\begin{equation}
V_{\rm 1-loop}(\q) = \frac{ 2\pi^2\alpha^2 }{ m^3 } 
\int \frac{ d^d k }{ (2\pi)^d } 
\frac{ [\si( \q-\k ),\sigma^i] }{ (\q-\k)^2 } \frac{ [\si'\k,\sigma'^i] }{\k^2 }.
\end{equation}

It induces the following correction to the ground state HFS 
($d$-dimensional integration over $\k$ is implicitly assumed below)
\begin{eqnarray}\label{E1loop}
\Delta_{\rm 1-loop} E_{\rm hfs} &=& - 2\frac{ d-1 }{ d }
       \frac{ \alpha^2 }{ m^3 }
       \left\langle \frac{ 4\pi( \p'-\k ) }{ (\p'-\k)^2 } 
       \frac{ 4\pi( \k -\p) }{ (\k-\p)^2 } \right\rangle \nonumber \\
       &=& - 2\frac{ d-1 }{ d }\frac{ \alpha^2 }{ m^3 }
       \left\langle \frac{ \p'^2 + \p^2 }{2} \frac{ 4\pi }{ (\p'-\k)^2 } 
       \frac{ 4\pi }{ (\k-\p)^2 }
       -  \frac{ 4\pi p'^i }{ (\p'-\k)^2 } 
       \frac{ 4\pi p^i }{ (\k-\p)^2 } \right\rangle \nonumber \\
       &=&  \frac{ \pi\alpha }{ m^4 } \left\{ 
       \frac{m^2\alpha^2}{3} \psi^2(0)
       - 8\frac{ d-1 }{ d }\left\langle  \frac{ \p'^2 \p^2 }{ \q^2 } 
       \right\rangle \right\}
       \nonumber \\
&=& 
-\frac{ 4\pi\alpha^3 }{ 3 m^2 } \psi^2(0)
                \left( \frac{ 1 }{\epsilon } - 4 \ln(m\alpha) - \frac{1}{3} \right). 
\label{Delta1}
\end{eqnarray}

\subsubsection{Breit Hamiltonian}
To complete the calculation of the soft scale
contributions to the HFS we have to consider 
the second iteration of the Breit Hamiltonian. 
It is obtained by including the effects of tree level Coulomb and
magnetic photon exchanges, as well as a correction to the kinetic
energy.
Using Eqs.~(\ref{eqa:coul}) and (\ref{eqa:magn}) we find
\begin{equation}\label{HBp}
U(\p',\p) = - \frac{ \p^4 }{ 4m^3 } (2\pi)^d\delta(\p'-\p)
              + \frac{ \pi\alpha }{ m^2 } 
              + \frac{ 4\pi\alpha }{ m^2 }\frac{(\p'\q)(\q\p)-(\p'\p) \q^2}{\q^4}
              - \frac{ \pi\alpha }{ 4m^2 }
                \frac{[\si\q,\sigma^i][\si'\q,\sigma'^i]}{\q^2}.
\end{equation}
In the position representation this Hamiltonian becomes
\begin{equation}\label{HBr}
U(\r,\p) = - \frac{ \p^4 }{ 4m^3 }
              + \frac{ d-1 }{ 4m } \left\{ \frac{ \p^2 }{ m },C(r) \right\} 
              + \frac{ d\pi\alpha }{ m^2 }\delta(\r) 
              - \frac{ 1 }{ 16m^2 }
                \left[ [\si\nabla,\sigma^i][\si'\nabla,\sigma'^i], C(r)\right],
\end{equation}
where 
\begin{equation}
C(r) \equiv - \frac{ \alpha\Gamma(d/2-1) }{ \pi^{d/2-1} r^{d-2} }
\end{equation}
is the $d$--dimensional Coulomb potential. 

\subsubsection{Second iteration of the Breit Hamiltonian: $S$-wave}
We consider first the contribution of the intermediate
$S$--states. The $S$--wave part of the Breit Hamiltonian
(\ref{HBr}) reads
\begin{equation}\label{BS}
U_S(\r,\p) = - \frac{ \p^4 }{ 4m^3 }
              + \frac{ d-1 }{ 4m } \left\{ \frac{ \p^2 }{ m },C(r) \right\} 
              + \frac{ d\pi\alpha }{ m^2 }\delta(\r) 
              - \frac{ \pi\alpha }{ 4dm^2 }
                [\sigma_i,\sigma_j][\sigma'_i,\sigma'_j] \delta(\r).
\end{equation}
It is convenient to divide up the calculation of the $U_S$
contribution to the HFS into two parts and consider the first and the
last two terms in Eq.~(\ref{BS}) separately.  We begin with the
latter, which we denote by $\Delta_{S1} E_{\rm hfs}$:
\begin{eqnarray}
\label{DES1}
\Delta_{S1} E_{\rm hfs} &=& 8\frac{(d-1)(3d-2)}{d^2} 
               \left( \frac{ \pi\alpha }{ m^2 } \right)^2
               \left\langle \delta(\r') \sum_m{\displaystyle'} \frac{
|m(\r') \rangle\langle m(\r) | }{ E-E_m } \delta(\r) \right\rangle \nonumber \\
&=& 8\frac{(d-1)(3d-2)}{d^2}
\left( \frac{ \pi\alpha }{ m^2 } \right)^2 \psi^2(0)
\sum_m{\displaystyle'} \frac{ |m(0)|^2 }{ E-E_m }.
\end{eqnarray}
In three dimensions the last sum is ill-defined due to ultraviolet
divergences in the zeroth and first terms of its expansion in
$\alpha$.  We denote these singular terms by $G_0(0,0)$ and
$G_1(0,0)$, respectively, and obtain
\begin{equation}
\Delta_{S1} E_{\rm hfs} = 8\frac{(d-1)(3d-2)}{d^2}
                  \left( \frac{ \pi\alpha }{ m^2 } \right)^2
                  \psi^2(0) 
                  \left[ - \frac{ 3m^2\alpha }{ 8\pi } 
                  + G_0(0,0) + G_1(0,0) \right].
\end{equation}
$G_0(0,0)$ and $G_1(0,0)$ are calculated in $d$ dimensions,
\begin{eqnarray}
G_0(0,0) &=& -  \int \frac{ {\rm d}^d p }{ (2\pi)^d
} \frac{ m }{\p^2  -m E } =
  \frac{m^2\alpha}{8\pi}, 
\nonumber \\
  G_1(0,0) &=& - \int \frac{
{\rm d}^d p' }{ (2\pi)^d } \int \frac{ {\rm d}^d p }
{ (2\pi)^d } \frac{ m }{\p'^2 - mE } 
\frac{ 4\pi\alpha }{ \q^2 } \frac{ m }{\p^2-m E } 
=- \frac{m^2\alpha}{16\pi}  
\left( \frac{ 1 }{\epsilon } - 4 \ln(m\alpha) + 2 \right),
\end{eqnarray}
and one finds
\begin{equation}
\Delta_{S1} E_{\rm hfs} = - \frac{ 7\pi\alpha^3 }{ 9 m^2 } \psi^2(0)
    \left( \frac{ 1 }{\epsilon } - 4 \ln(m\alpha) + \frac{115}{21} \right).
\label{DeltaS1}
\end{equation}

The contribution of the first two terms in Eq.~(\ref{BS}) is
calculated in the following way. We first write them as
\begin{equation}\label{S2}
-\frac{ \p^4 }{ 4m^3 } + \frac{ d-1 }{ 4m } \left\{ \frac{ \p^2 }{ m },
C(r) \right\} = - \frac{1}{ 4m } 
\left[ H^2 - d\left\{ H, C(r) \right\} + (2d-1) C^2(r) \right],
\end{equation}
where $H = \p^2/m + C(r)$ is the leading order Hamiltonian.
Correction to the HFS induced by Eq.~(\ref{S2}) reads
\begin{equation}\label{DES2}
\Delta_{S2} E_{\rm hfs} = \frac{d-1}{d} \frac{ \pi\alpha }{ m^3 }
 \left\langle  
 \left[ d\left\{ H, C(r) \right\} -(2d-1) C^2(r) \right] 
       G(\r,\r') \delta(\r') + {\rm H.c.}  
 \right\rangle.
\end{equation}
We introduced here the reduced Green function
\begin{equation}
G(\r,\r') = \sum_m{\displaystyle'} \frac{ |m(\r) \rangle\langle m(\r') | }{
E-E_m },
\end{equation}
which satisfies the equation
$
(H - E)G(\r,\r') = \psi(r)\psi(r') - \delta(\r-\r').
$
Using obvious short-hand notations one can rewrite Eq.~(\ref{DES2}) as
follows: 
\begin{eqnarray}
\Delta_{S2} E_{\rm hfs} &=& - \frac{ 2\pi\alpha }{ 3m^3 }
                    \left\langle \left( 6 \alpha E \frac{1}{r} G 
                    + 3 \frac{ \alpha }{r}\psi(r)\psi(r')
                    + 5 \frac{ \alpha^2 }{ r^2 } ( G - G_0 )
                    \right.\right. \nonumber 
\\
 && \qquad \qquad\qquad\qquad \left.\left.  
 + \frac{3}{2} \frac{(d-1)(2d-1)}{d}  C^2(r) G_0 \right) \delta(\r') 
 + {\rm H.c.} \right\rangle,
\label{57}
\end{eqnarray}
We dropped massless tadpoles and separated the
contribution of $G_0$, which is the only one we have to calculate
keeping $d \neq 3$.  We find
\begin{eqnarray}
\left\langle C^2 G_0 \delta(\r') + {\rm H.c.} \right\rangle &=&
\frac{ 8\pi\alpha }{ m } \psi^2(0) G_1(0,0),
\nonumber \\
\left\langle \frac{\alpha}{r} G \delta(\r') + {\rm H.c.} \right\rangle &=& 
                       -\alpha \partial_{\alpha} \left\langle \delta(\r) \right\rangle = 
                       - 3\psi^2(0),
\nonumber  \\
\left\langle \frac{ \alpha }{ r } \right\rangle &=& -2E,
\nonumber  \\
\left\langle \frac{ 1 }{ r^2 } ( G - G_0 ) \delta(\r') + {\rm H.c.} \right\rangle &=& -4m\psi^2(0).
\end{eqnarray}
To obtain the last line we used the following equation:
\begin{equation}
G(r,0) - G_0(r,0) = \frac{m^2\alpha}{4\pi} e^{-\gamma r}
        \left( \ln (2\gamma r) + \gamma_E - \frac{5}{2} + \gamma r \right),
\end{equation}
where $\gamma=m\alpha/2$. 
>From Eq.~(\ref{57}) we now find
\begin{equation}
\Delta_{S2} E_{\rm hfs} = \frac{ 5\pi\alpha^3 }{ 3 m^2 } \psi^2(0)
                \left( \frac{ 1 }{\epsilon } - 4 \ln(m\alpha) + \frac{88}{15} \right).
\label{DeltaS2}
\end{equation}

The sum of $\Delta_{S1}E_{\rm hfs}$ and $\Delta_{S2}E_{\rm hfs}$ gives
the final result for the correction to the ground state HFS induced by
the second iteration of the $S$-wave Breit Hamiltonian:
\begin{equation}
\Delta_{S}E_{\rm hfs} = \frac{ 8\pi\alpha^3 }{ 9 m^2 } \psi^2(0)
                \left( \frac{ 1 }{\epsilon } - 4 \ln(m\alpha) + \frac{149}{24} \right).
\label{Swavehfs}
\end{equation}

\subsubsection{Second iteration of the Breit Hamiltonian: $D$-wave}
Because of the last term in Eq.~(\ref{HBp}) Breit Hamiltonian has
non-vanishing matrix elements with $|\Delta L| = 2$. In our case this
causes virtual transitions from the triplet $S$--state into
$D$--states (transitions from the singlet state are forbidden by the
total angular momentum conservation).  Again, power counting shows that
only the zeroth and the first order terms in the Green function
expansion in $\alpha$ diverge in three dimensions.  We first compute
the remaining, higher order terms, which are finite for $d=3$.

The sum of those higher order terms can be written as
\begin{equation}\label{D}
\Delta_D^{\rm h-o} E_{\rm hfs} =  \left\langle U_D G_0 C G C G_0 U_D
\right\rangle, 
\end{equation}
where
\begin{equation}
U_D = \frac{\alpha}{4m^2}
\frac{ 3(\vec{\sigma}\vec{n})(\vec{\sigma}'\vec{n}) -
\mbox{$\vec{\sigma}\vec{\sigma}'$} }{ r^3 } 
\end{equation}
is the $|\Delta L| = 2$ part of the Breit Hamiltonian in three
dimensions, $G$ and $G_0$ are defined in the previous Section,
and $C=-\alpha/r$ is the Coulomb potential.

The correction to the ground--state wave function,
\begin{equation}
\delta_0\psi(r) = G_0 U_D \psi(r),
\end{equation}
which appears in Eq.~(\ref{D}), satisfies an inhomogeneous
Schr\"odinger equation: 
\begin{equation}
\left( E - \frac{\p^2}{m} \right) \delta_0\psi(r) =  U_D \psi(r).
\end{equation}
Solving this equation for $\delta_0\psi(r)$ we obtain
\begin{equation}
\Delta_D^{\rm h-o} E_{\rm hfs} = 8 \delta_{S1} \left
                       ( \frac{\alpha^2}{24} \right)^2 
                       \left\langle \frac{1}{mr^2} G_D(r,r_1)
                       \frac{1}{mr_1^2} \right\rangle, 
\label{72}
\end{equation}
where $G_D(r,r_1)$ is the $D$-wave part of the Green function $G$, and 
the factor $8\delta_{S1}$ arises from
\begin{equation}
\left\langle
\left
[
\mbox{$\vec{\sigma}\vec{\sigma}'$}-3(\vec{\sigma}\vec{n})(\vec{\sigma}'\vec{n})
\right]^2 
\right\rangle = \left\langle
3+4\mbox{$\vec{\sigma}\vec{\sigma}'$}+(\mbox{$\vec{\sigma}\vec{\sigma}'$})^2
\right\rangle = 8\delta_{S1}. 
\end{equation}
To calculate the matrix element in Eq.~(\ref{72}) we note that
\begin{equation}
\frac{1}{mr^2} = \frac{1}{6} \left( H_D - H \right),
\end{equation}
where $H_D(H)$ is the radial Hamiltonian for $D(S)$-states. 
Using equations of motion for both the Green function and 
the wave function in Eq.~(\ref{72}) one finds
\begin{equation}\label{deltaD}
\Delta_D^{\rm h-o} E_{\rm hfs} = 8 \left( \frac{\alpha^2}{24} \right)^2
                       \left\langle - \frac{1}{6mr^2}  \right\rangle
                       = - \frac{ \pi\alpha^3\psi^2(0) }{ 108m^2 },
\end{equation}
in agreement with \cite{Bodwin:1978ut}.

To complete the calculation of the $D$-wave contribution
we have to consider the 
zeroth and first order terms in the $\alpha$
expansion of the Green function,
\begin{eqnarray}
\Delta_{D0} E_{\rm hfs} &=& \left\langle U_D G_0 U_D \right\rangle,
                 \label{D0r}
\\
\Delta_{D1} E_{\rm hfs} &=& \left\langle U_D G_1 U_D \right\rangle. \label{D1r}
\end{eqnarray}
The perturbation $U_D(\p',\p)$ is extracted from Breit Hamiltonian,
Eq.~(\ref{HBp}), and reads
\begin{equation}
U_D(\p',\p) = \frac{ \pi\alpha }{ 4m^2 }
           \left( \frac{ [\sigma_i,\sigma_j][\sigma'_i,\sigma'_j] }{ d }
              - \frac{[\si\q,\sigma_i][\si'\q,\sigma'_i]}{\q^2} \right).
\end{equation}
The average is taken over the $d$--dimensional wave function.
Calculating the trace using the triplet wave function we obtain
\begin{eqnarray}
&&\left\langle \left( \frac{ [\sigma_i,\sigma_j][\sigma'_i,\sigma'_j] }{ d }
              - \frac{[\si\q',\sigma^i][\si'\q',\sigma'^i]}{\q'^2} \right)
    \left( \frac{ [\sigma_k,\sigma_l][\sigma'_k,\sigma'_l] }{ d }
              - \frac{[\si\q,\sigma^k][\si'\q,\sigma'^k]}{\q^2} \right)
    \right\rangle  \nonumber \\
&&\qquad\qquad\qquad
  = \frac{4(d-2)^2}{d}\left\langle B_{ij}(\q') B_{ij}(\q) \right\rangle, \qquad
B_{ij}(\q) \equiv 4\pi \left( \frac{  q_i q_j }{ \q^2 }
                      - \frac{\delta_{ij}}{ d } \right).
\end{eqnarray}
Therefore
\begin{eqnarray}
\Delta_{D0} E_{\rm hfs} &=& - \frac{ \alpha(d-2)^2 }{ 4m^4d }
                    \left\langle B_{ij}(\p'-\k)g(k)B_{ij}(\k-\p) \right\rangle, \label{D02}
                 \\
\Delta_{D1} E_{\rm hfs} &=& - \frac{ \alpha(d-2)^2 }{ 4m^4d }
                    \left\langle B_{ij}(\p'-\k')g(k')
                    \frac{4\pi}{(\k'-\k)^2}g(k)B_{ij}(\k-\p) \right\rangle, \label{D12}
\end{eqnarray}
where
\begin{equation}
g(k) = \frac{ 2\gamma }{ \k^2 + \gamma^2 },
\end{equation}
and $d$--dimensional integrations over $\k$ in (\ref{D02}) and over
$\k,\k'$ in (\ref{D12}) are understood.  Some details of the
integrations in Eqs.~(\ref{D02},\ref{D12}) are given
in Appendix \ref{app:int}.  
Adding the higher-order
effects found in Eq.~(\ref{deltaD}) we obtain the complete
$D$-wave contributions to HFS 
\begin{eqnarray}
\Delta_{D} E_{\rm hfs} &=&
\Delta_{D0} E_{\rm hfs} + \Delta_{D1} E_{\rm hfs} 
+\Delta_D^{\rm h-o} E_{\rm hfs}
\nonumber \\
&=&
\frac{ 5\pi\alpha^3 }{ 72 m^2 } \psi^2(0)
  \left( \frac{ 1 }{\epsilon } - 4 \ln(m\alpha) - \frac{19}{5} \right).
\label{DeltaD2}
\end{eqnarray}

\subsection{Hard scale contribution}
\label{sec:hard}

Another contribution to the HFS arises from virtual momenta scales of
the order of the electron mass. It can be calculated by considering
the on--shell $e^+e^-$ scattering amplitude with an exchange of three
photons in the $t$-channel (see Fig.~\ref{fig1}) exactly at the
threshold, i.e.~for zero relative velocity of the incoming electron
and positron, in dimensional regularization.  The use of the
dimensional regularization  brings in essential simplifications, 
since almost any other
regularization would bring in power-like divergences and hence require
additional subtractions.  This so-called hard scale contribution gives
rise to four-fermion operators in the low-scale Lagrangian or,
equivalently, to the $\delta(r)$ terms in the effective quantum
mechanical Hamiltonian.

Technically, this calculation is similar to the derivation of 
the matching coefficient of the vector quark-antiquark current in QCD
and its NRQCD counterpart, described e.g. in \cite{threshold,BSS}.
Here we outline the main steps of this calculation.

An arbitrary Feynman integral which contributes to the hard scale part of
the calculation can be written as
\begin{equation}
I(a_1,...a_9) = \int \frac {{\rm d}^D k_1}{(2\pi)^D} \frac {{\rm d}^D k_2}{(2\pi)^D}
{1\over S_1^{a_1}S_2^{a_2}S_3^{a_3}S_4^{a_4}S_5^{a_5}S_6^{a_6}S_7^{a_7}
S_8^{a_8} S_9^{a_9}},
\end{equation}
where
\begin{eqnarray}
&&
S_1 = {k_1^2},\qquad   
S_2={k_2^2},\qquad   
S_3={(k_1-k_2)^2}, \qquad    
S_4 = {k_1^2+2pk_1},
\nonumber \\
&&
S_5= {k_2^2+2pk_2},\qquad
S_6 = {k_1^2-2pk_1},\qquad    
S_7 = {k_2^2-2pk_2},
\nonumber \\
&&
S_8 = {(k_1-k_2)^2+2p(k_1-k_2)},\qquad
S_9 = {(k_1-k_2)^2-2p(k_1-k_2)},
\end{eqnarray}
and $a_1,\ldots,a_9$ are integers.  In practice we encounter diagrams
with 
only at most 6 different propagators, so that at least 3 exponents
$a_i$ are zero.  Applying the integration by parts technique
\cite{che81} to an integral $I(\{a_i\})$, one obtains a set of relations
among integrals with various values of indices $\{a_i\}$.  Using these
relations one can express any $I(\{a_i\})$ in terms of a few master
integrals.  This is most easily done using symbolic manipulation
programs.

The result for the hard scale recoil corrections (Fig.~\ref{fig1})
to the HFS reads \cite{Czarnecki:1998zv}
\begin{equation}
\Delta_{\rm hard} E_{\rm hfs} = \frac{ \pi\alpha^3 }{ 3 m^2 } \psi^2(0)
                \left( - \frac{ 1 }{\epsilon } + 4 \ln m 
               - \frac{51\zeta(3)}{\pi^2} + 
               \frac{10}{\pi^2} - 6\ln 2 \right).
\label{hard}
\end{equation}

\subsection{HFS for excited \boldmath $S$ states}
\label{sec:nhfs}

The result for the HFS of the ground state can be used to obtain the
HFS for an arbitrary excited state.  The non--trivial dependence on
the principal quantum number
$n$ arises only from the soft scale contributions.  Therefore, one has
to repeat the quantum mechanical calculation of the non-relativistic
part using any convenient regularization (we use a cut-off at $1/m \ll
r_0\ll
1/m\alpha$)  and compare the result with
the known formula for $n=1$, Eq.~(\ref {Ehfsrecground}).  One finds
\begin{equation}
\Delta_{\rm rec} E_{\rm hfs}(n) = \frac {m \alpha^6}{n^3} \left [ [div] 
-\frac {1}{6} \left ( \ln \frac {\alpha}{n} +\Psi (n) + \gamma_E \right ) 
+ \frac {7}{12 n} - \frac {1}{2n^2}
\right ],
\end{equation}
The quantity $[div]$ in the above equation stands
for the unknown and $n$-independent constant,
easily
determined by  requiring that for $n$=1  Eq.~(\ref {Ehfsrecground})
is reproduced.  We then obtain the final result for the 
recoil corrections to the HFS splitting
for an arbitrary $nS$ state, Eq.~(\ref{Ehfsrecn}).

\section{Spin-averaged energy levels}
\label{sec:lev}
To obtain ${\cal O}(m\alpha^6)$ corrections to the triplet and singlet
energy levels separately, we have to calculate $E_{\rm aver}(n)$
(cf. Eq.~(\ref{param})).  An appropriate formula for this calculation
is
$$
E_{\rm aver} = \frac {3\, E_{\rm triplet} + E_{\rm
singlet}}{4} \to \frac {d\; E_{d\rm -plet} + E_{\rm singlet}}{d+1}.
$$
It is known \cite{Fell1,Fell2,KMYlog} that the recoil corrections 
$E_{\rm aver}$ do not
contain $\ln(\alpha)$ at the order $m\alpha^6$.  In dimensional
regularization this means that the hard-scale and soft-scale
contributions are separately finite.\footnote{Because of power-like
singularities, this is not necessarily the case in other
regularization schemes.}

Conceptually, determination of $E_{\rm aver}$ is very similar to the
calculation of the HFS discussed above in detail.  The only difference
is that several new operators appear, which contribute to 
$E_{\rm aver}$ but not to the HFS.

\subsection{The ground state average energy shift}
\begin{figure} 
\hspace*{-2mm}
\begin{minipage}{16.cm}
\vspace*{3mm}
\[
\hspace*{-5mm}
\mbox{ 
\begin{tabular}{ccc}
\psfig{figure=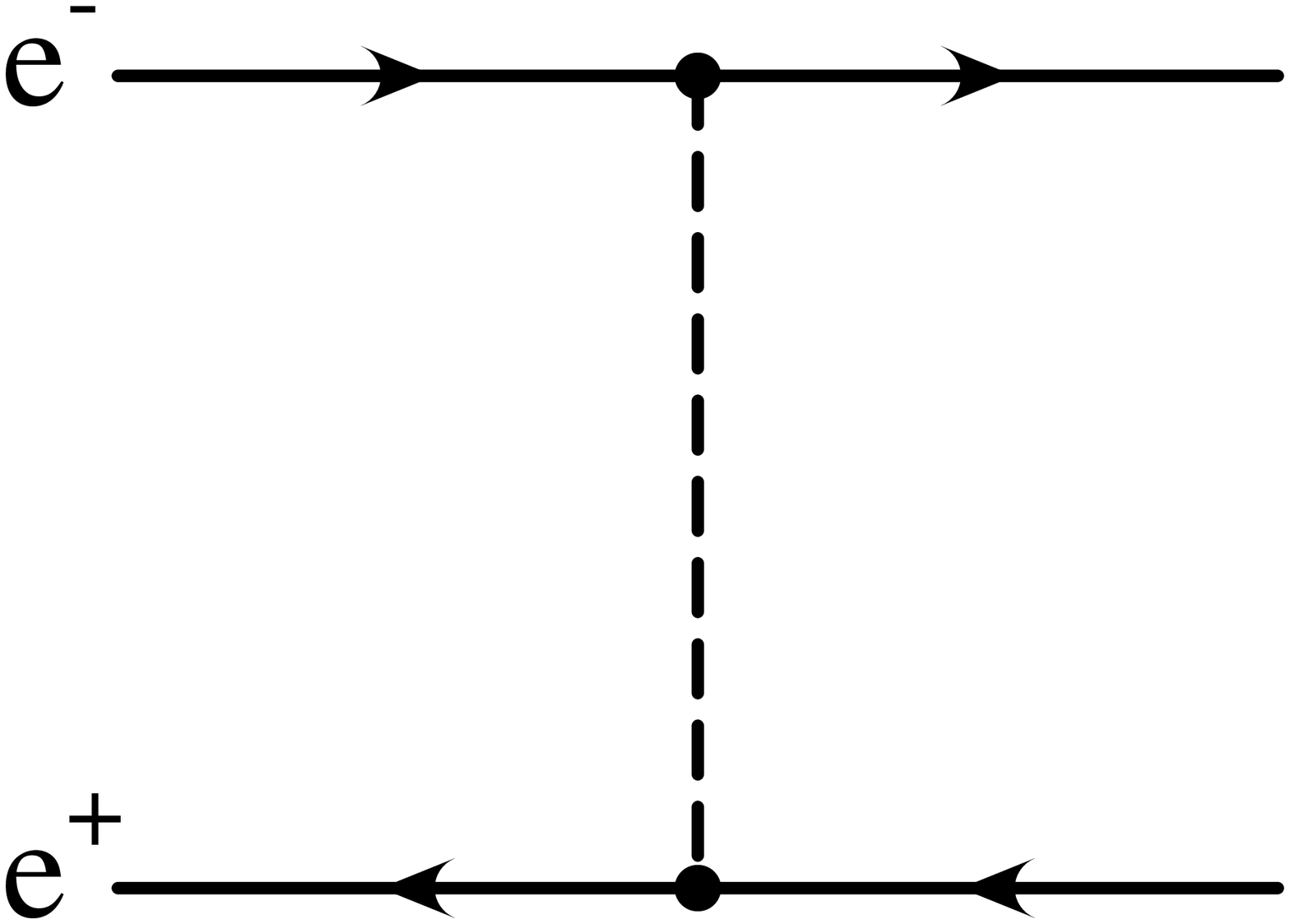,width=30mm,bbllx=72pt,bblly=291pt,%
bburx=544pt,bbury=530pt} 
& \hspace*{10mm}
\psfig{figure=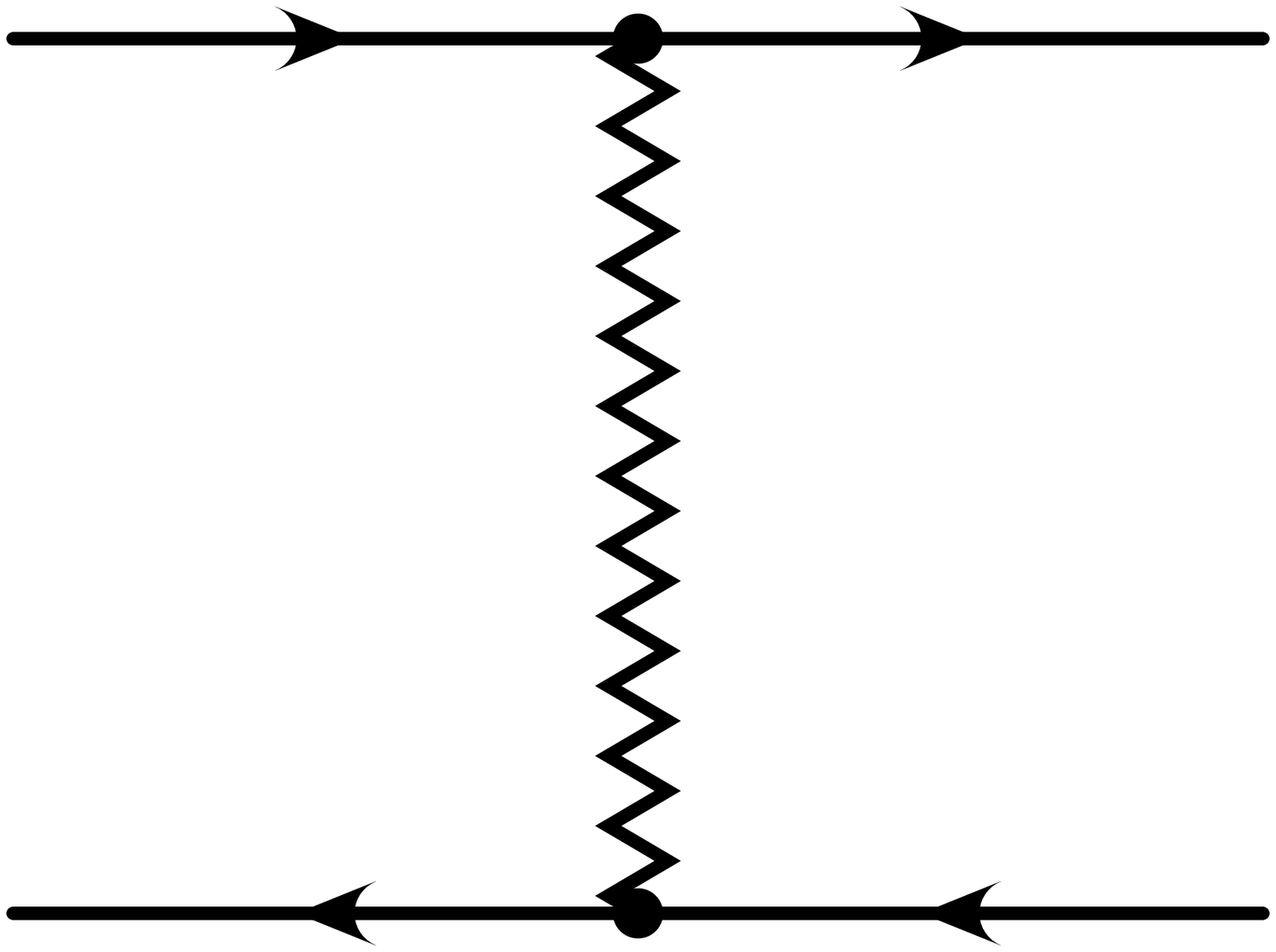,width=30mm,bbllx=72pt,bblly=291pt,%
bburx=544pt,bbury=530pt} 
& \hspace*{10mm}
\psfig{figure=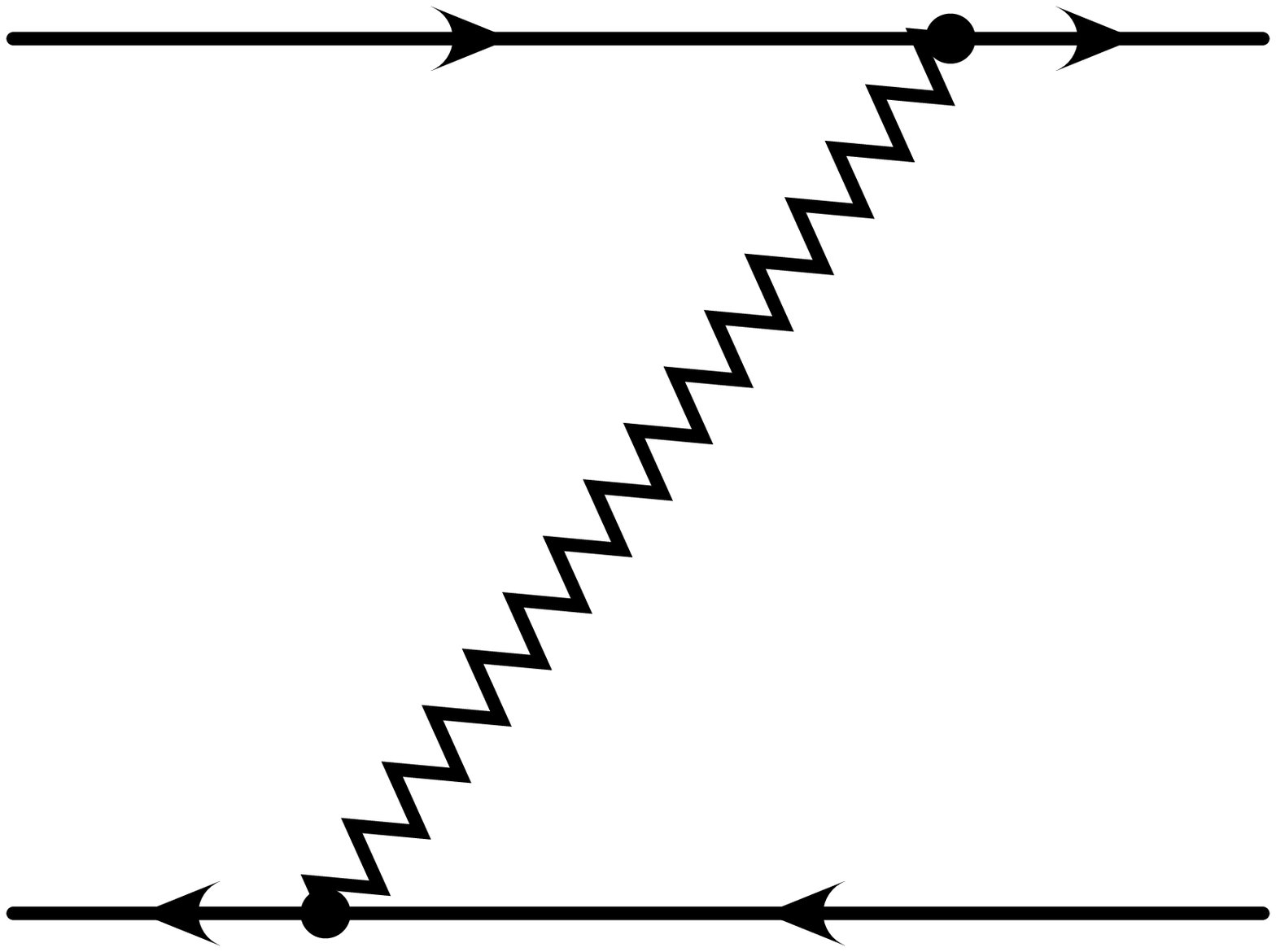,width=30mm,bbllx=72pt,bblly=291pt,%
bburx=544pt,bbury=530pt} 
\\[7mm]
(a) & \hspace*{11mm}(b) &\hspace*{10mm} (c)
\\[9mm]
\psfig{figure=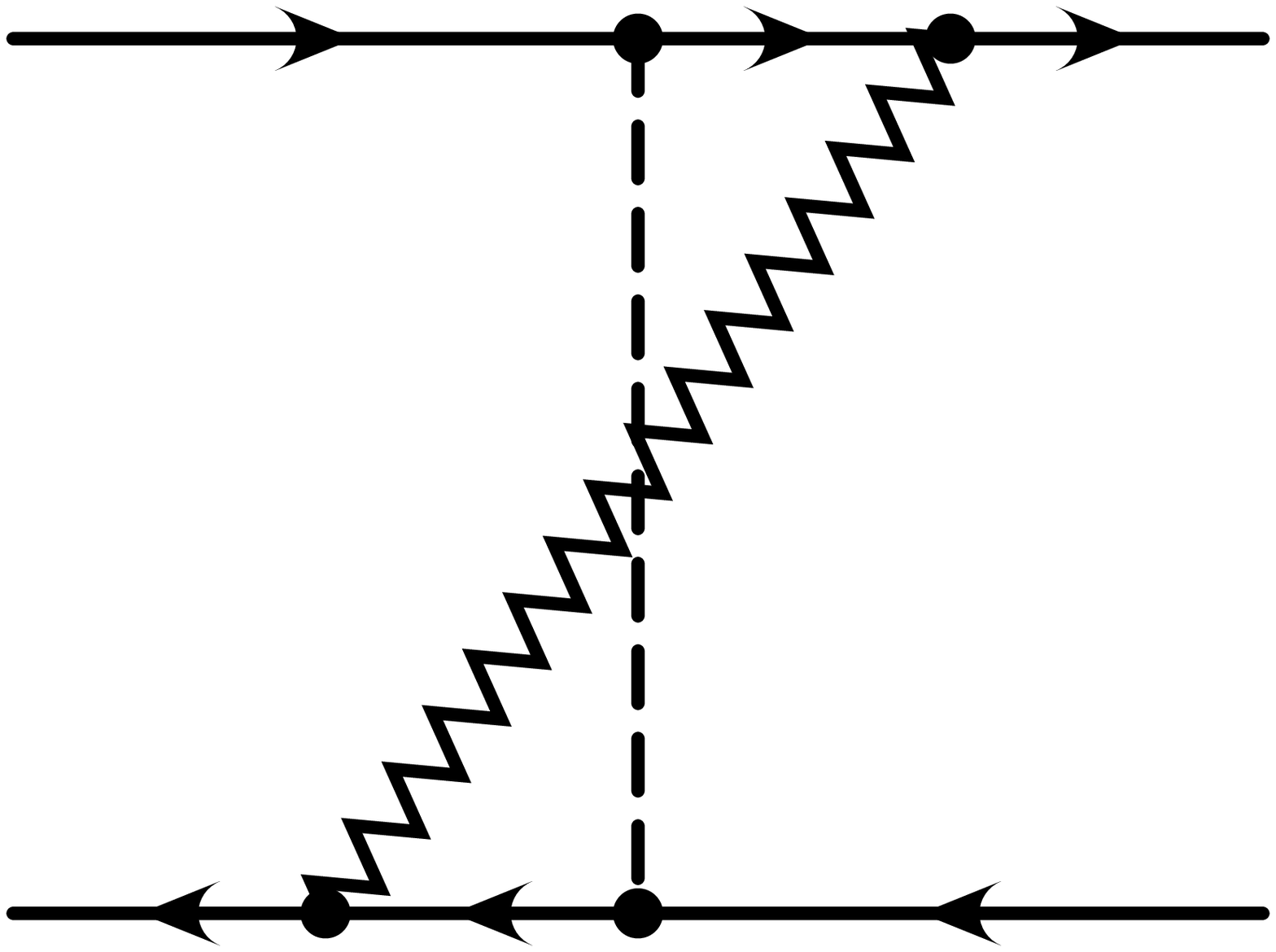,width=30mm,bbllx=72pt,bblly=291pt,%
bburx=544pt,bbury=530pt} 
& \hspace*{10mm}
\psfig{figure=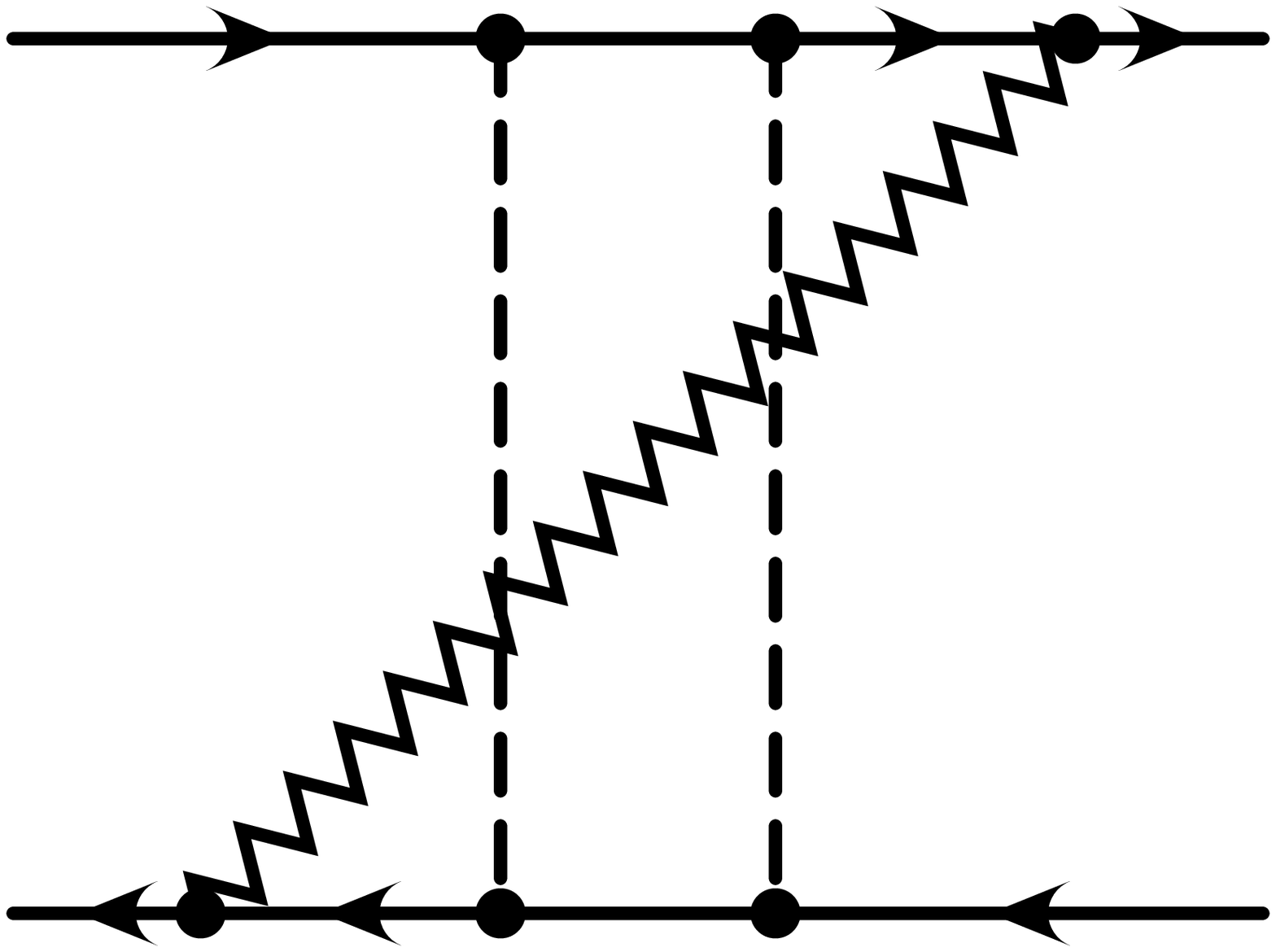,width=30mm,bbllx=72pt,bblly=291pt,%
bburx=544pt,bbury=530pt} 
& \hspace*{10mm}
\psfig{figure=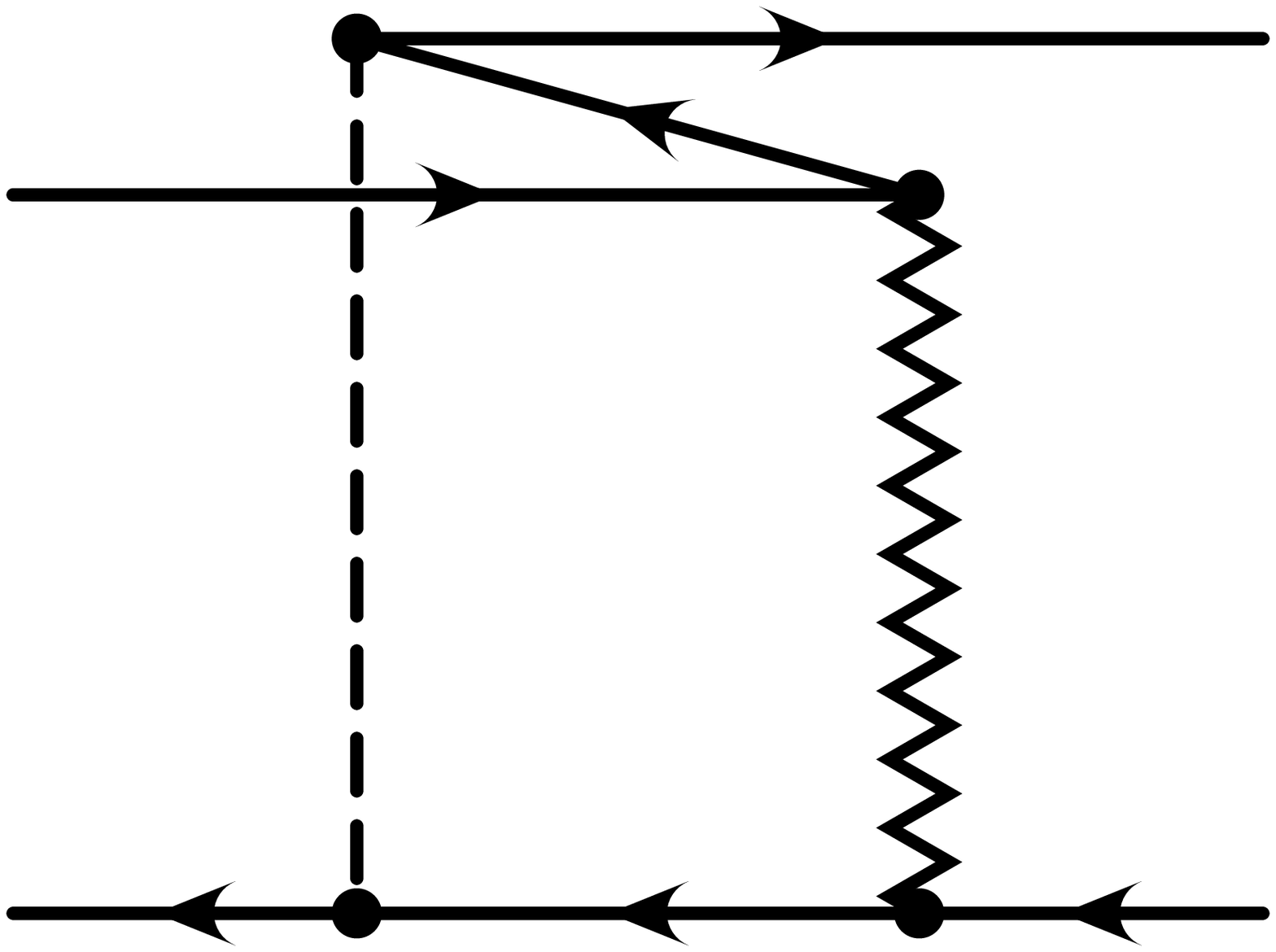,width=30mm,bbllx=72pt,bblly=291pt,%
bburx=544pt,bbury=530pt} 
\\[7mm]
(d) & \hspace*{11mm}(e) &\hspace*{10mm} (f)
\\[9mm]
\psfig{figure=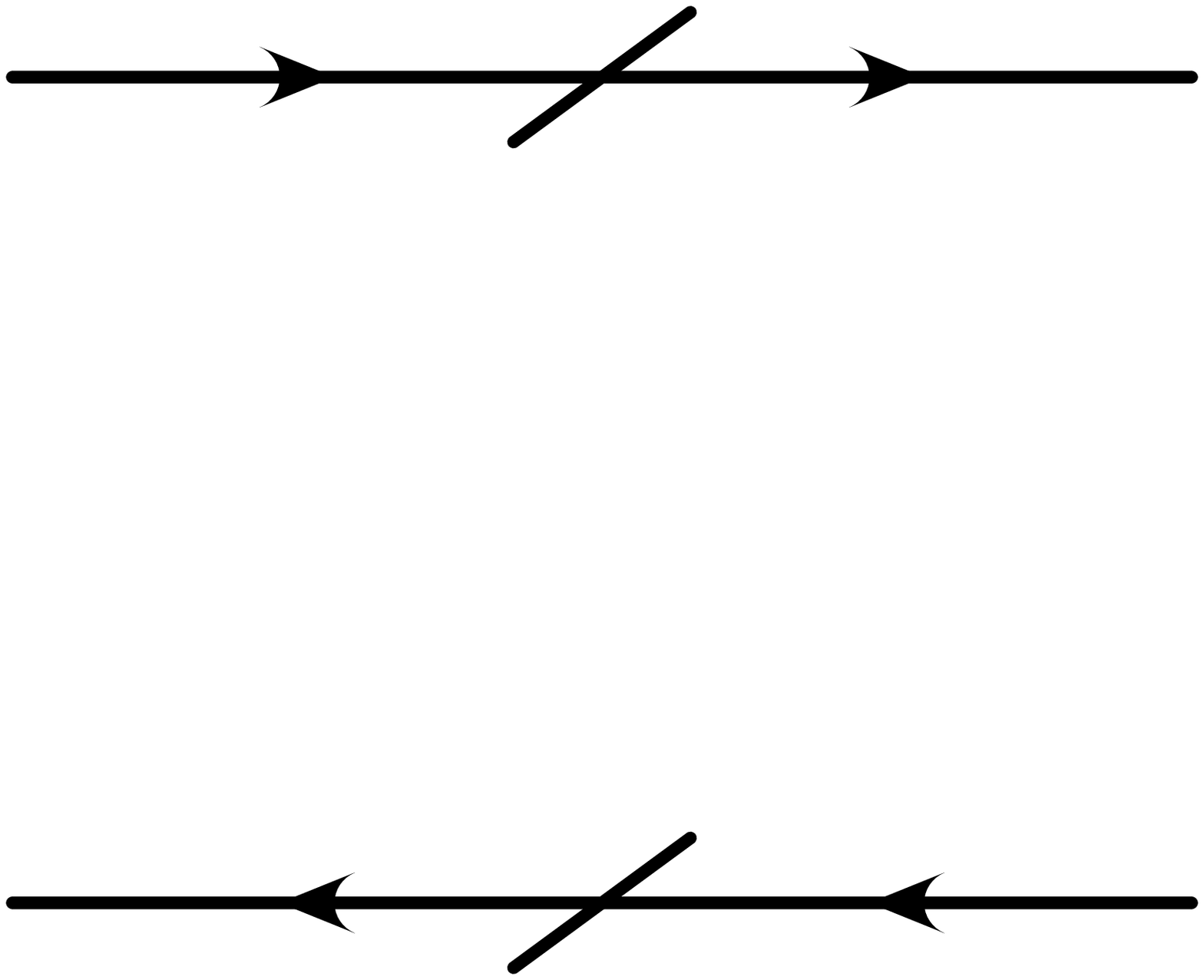,width=30mm,bbllx=72pt,bblly=291pt,%
bburx=544pt,bbury=530pt} 
& \hspace*{10mm}
\psfig{figure=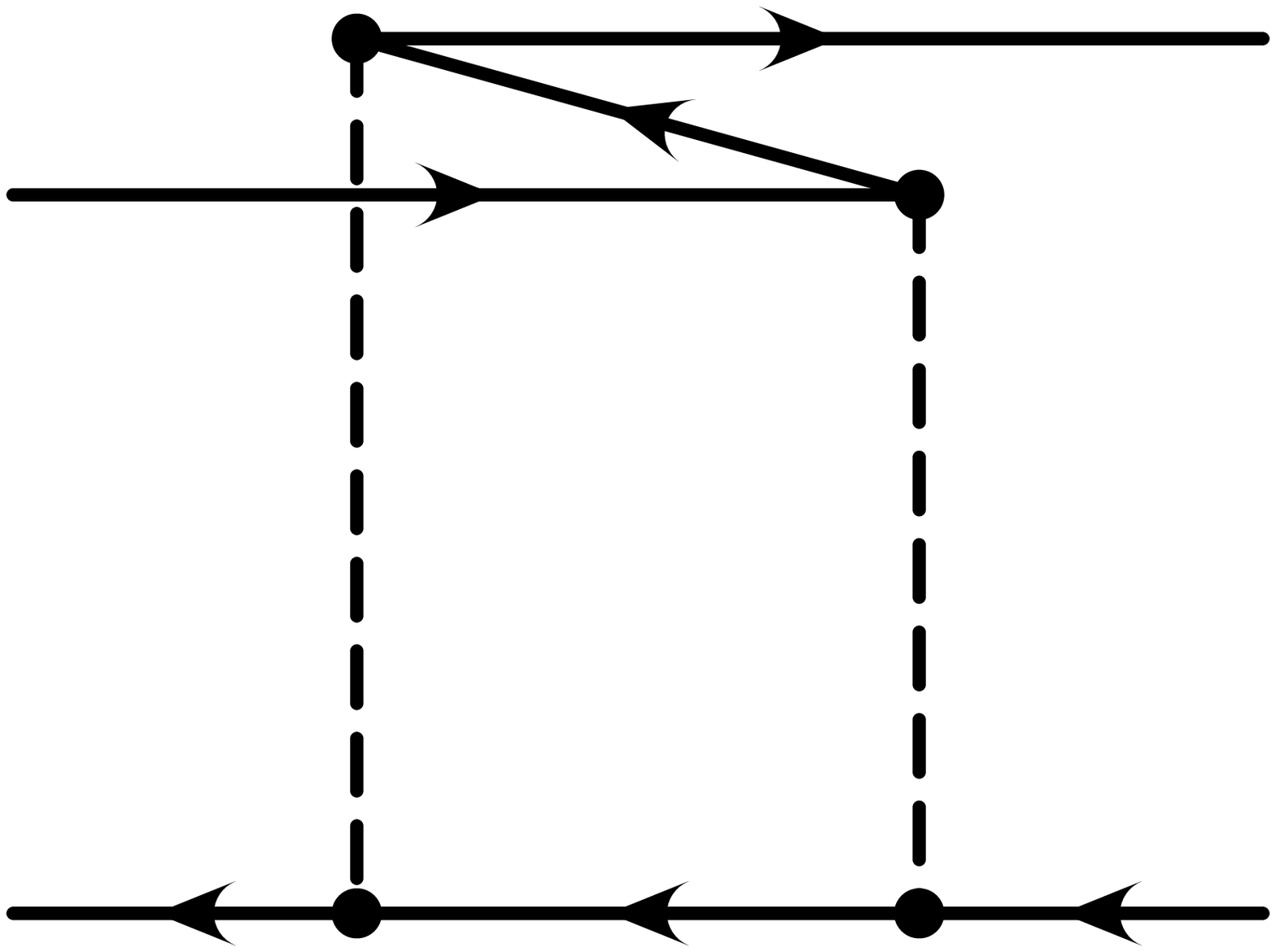,width=30mm,bbllx=72pt,bblly=291pt,%
bburx=544pt,bbury=530pt} 
& \hspace*{10mm}
\psfig{figure=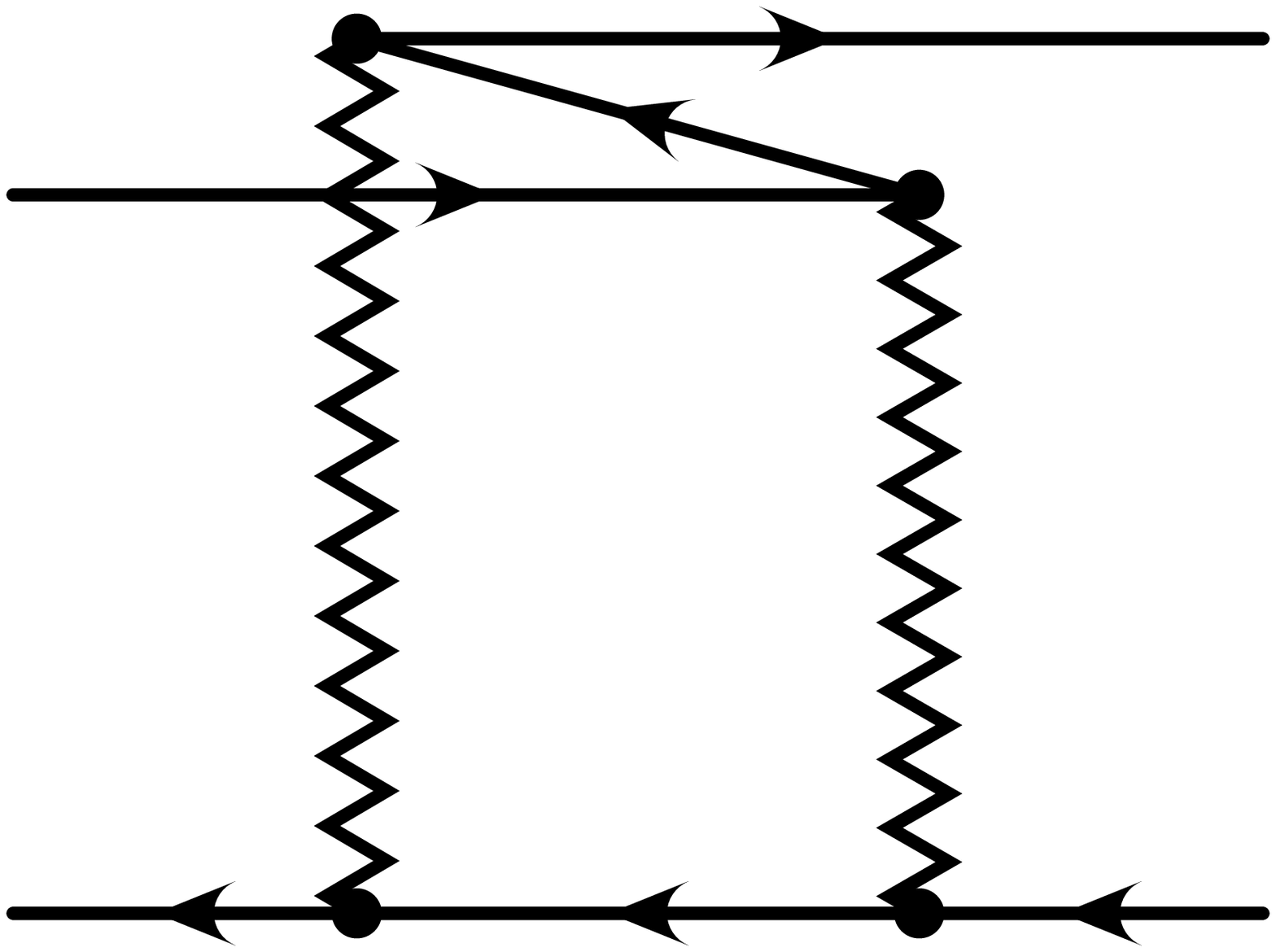,width=30mm,bbllx=72pt,bblly=291pt,%
bburx=544pt,bbury=530pt} 
\\[7mm]
(g) & \hspace*{11mm}(h) &\hspace*{10mm} (i)
\\[2mm]
\end{tabular}
}
\]
\end{minipage}
\caption{Non-relativistic corrections to HFS and spin-averaged energy
levels: (a,b) Coulomb and magnetic photon exchange; (c,d,e)
retardation effects; (f) mixed Coulomb-magnetic exchange; (g)
relativistic correction to the dispersion law; (h,i) double Coulomb
and magnetic exchange.}
\label{fig:soft}
\end{figure}

\begin{figure} 
\hspace*{-2mm}
\begin{minipage}{16.cm}
\vspace*{3mm}
\[
\hspace*{-5mm}
\mbox{ 
\begin{tabular}{cc}
\psfig{figure=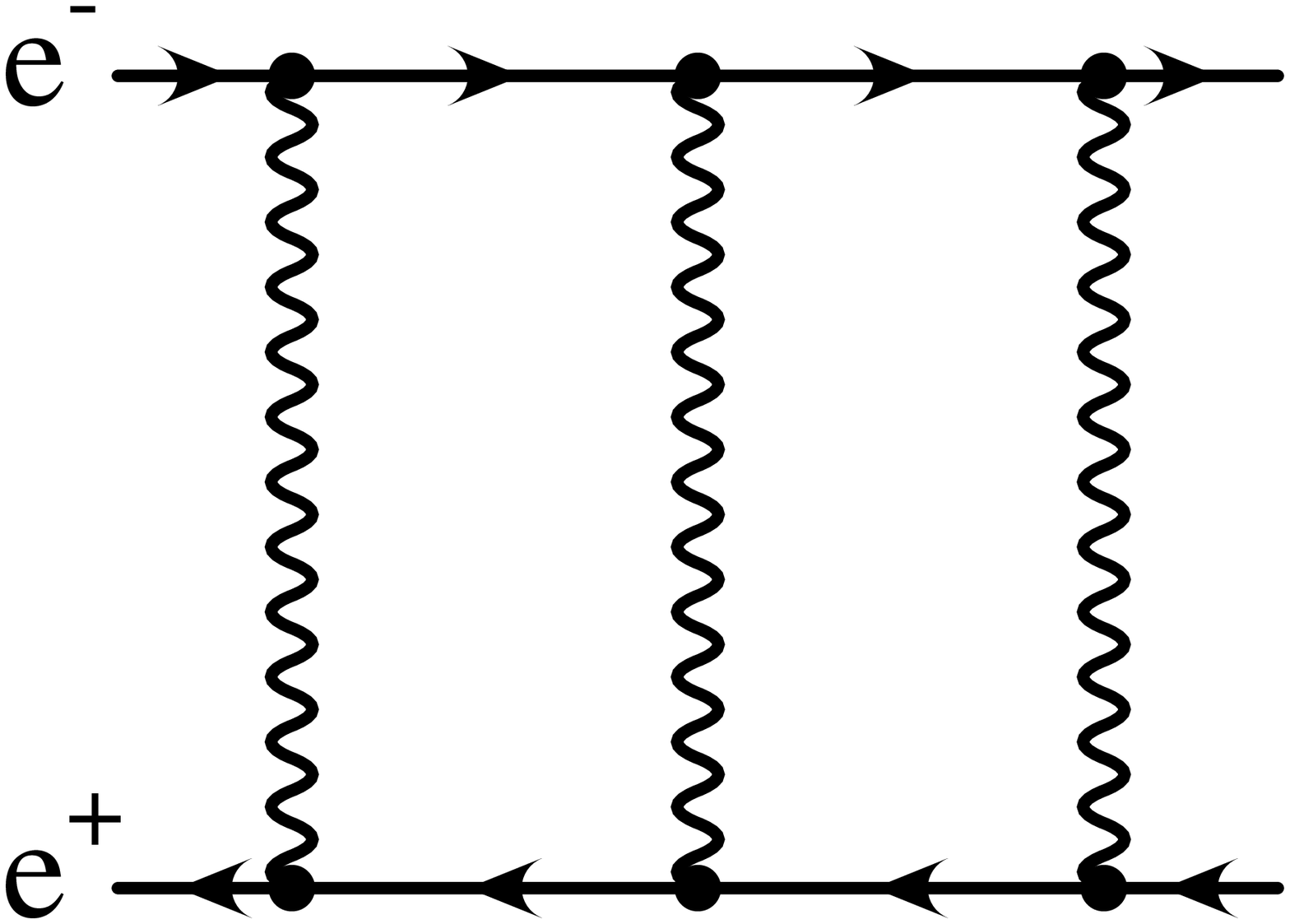,width=30mm,bbllx=72pt,bblly=291pt,%
bburx=544pt,bbury=530pt} 
& \hspace*{10mm}
\psfig{figure=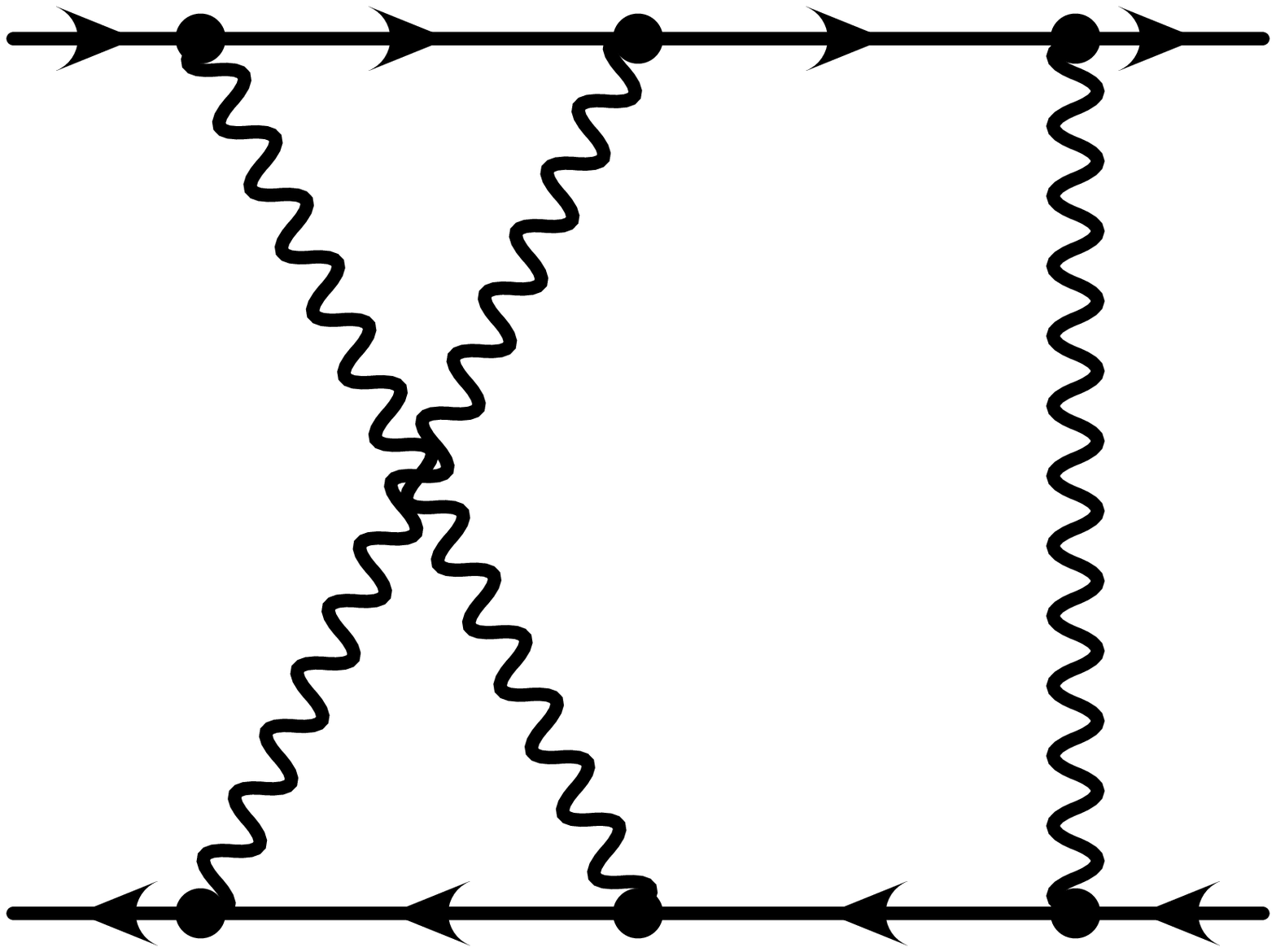,width=30mm,bbllx=72pt,bblly=291pt,%
bburx=544pt,bbury=530pt} 
\\[12mm]
\psfig{figure=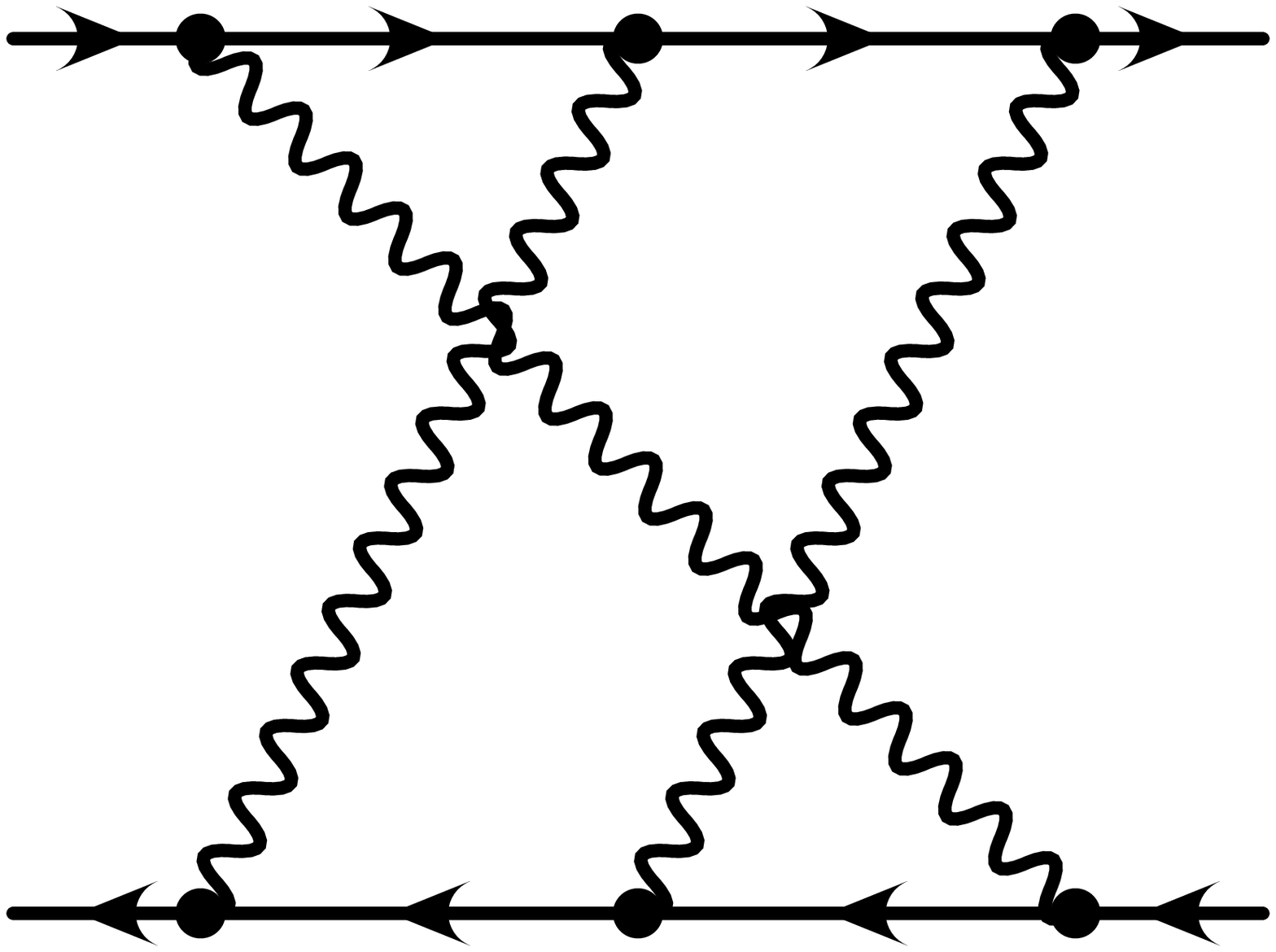,width=30mm,bbllx=72pt,bblly=291pt,%
bburx=544pt,bbury=530pt} 
& \hspace*{10mm}
\psfig{figure=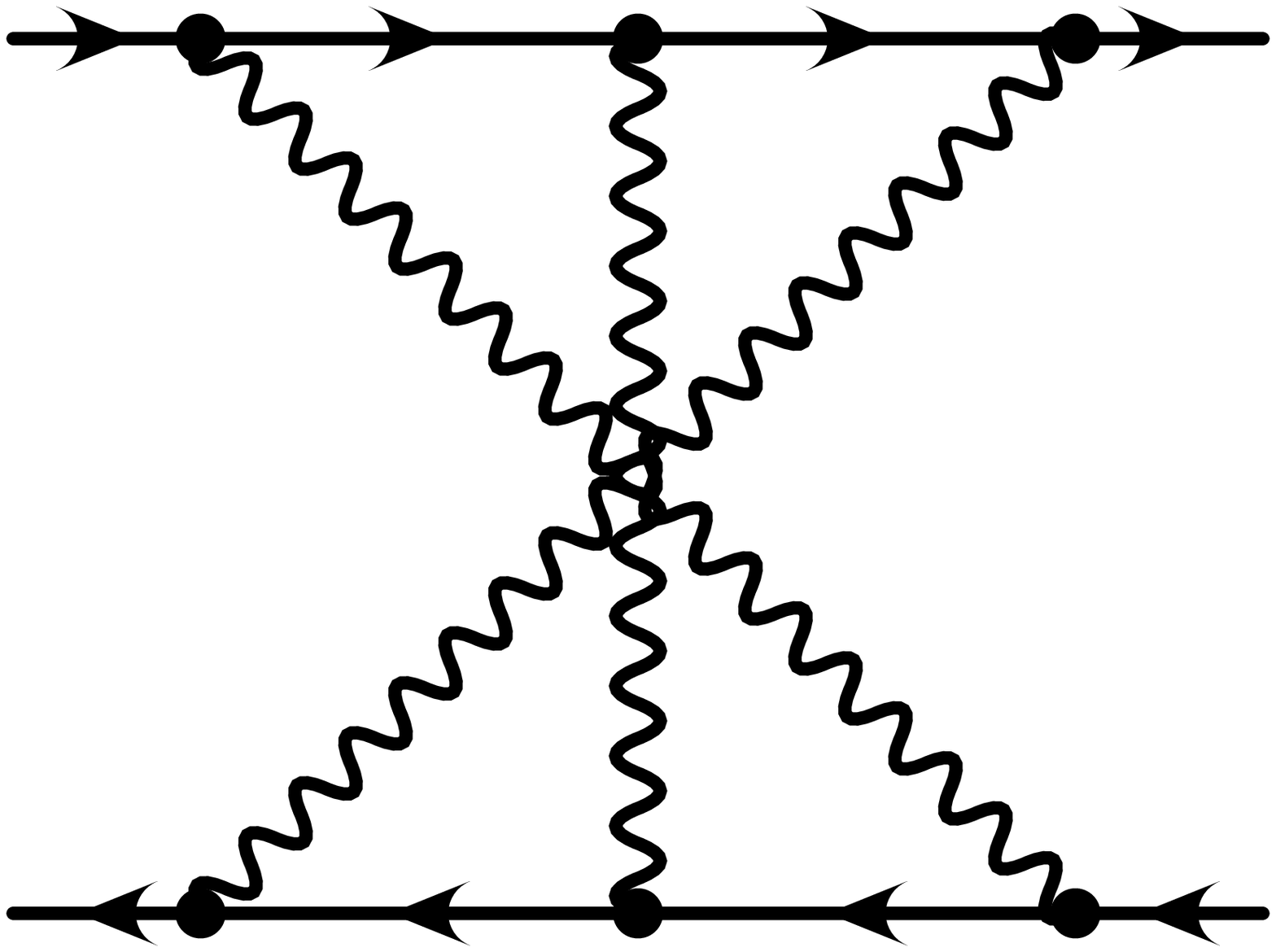,width=30mm,bbllx=72pt,bblly=291pt,%
bburx=544pt,bbury=530pt} 
\end{tabular}
}
\]\vspace*{6mm}
\end{minipage}
\caption{Feynman diagrams representing pure recoil corrections to
positronium HFS and spin-averaged energy levels.  Wiggly lines denote
photons in Feynman gauge.}
\label{fig1}
\end{figure}

We begin with the correction to $E_{\rm aver}$ induced by the
relativistic corrections to the dispersion law, $\omega_{p} =
\sqrt{\p^2+m^2}$ (Fig.~\ref{fig:soft}(g)). Expanding $\omega_p$ in
$|\p|/m$, we obtain:
\begin{equation}
\omega_{p}=m+\frac {\p^2}{2m}-\frac {\p^4}{8m^3}+\frac {\p^6}{16m^5}+\ldots.
\end{equation}
The last term induces a correction of the appropriate order:
\begin{equation}
\Delta_{\rm disp} E_{\rm aver} =
\frac {1}{8m^5} \left\langle  \p^6 (2\pi)^d \delta^{(d)} (\p - \p ')  \right\rangle = 
-\frac{3}{64}\frac{\pi\alpha^3}{m^2}\psi^2(0).
\label{averdips}
\end{equation}

The ${\cal O}(v^4)$ spin-independent part of the
tree level Coulomb exchange amplitude (cf.~Eq.~(\ref{AC}) and
Fig.~\ref{fig:soft}(a); we neglect terms odd in $\p$, whose average
vanishes in an $S$-state), 
\begin{equation}
\overline V_C(\p',\p) = - \frac{\pi\alpha}{16m^4} \left( 
7\left(\p^2 + \p'^2\right)
+ \frac{5(\p^2 - \p'^2)^2}{\q^2} \right),
\end{equation}
gives rise to the following correction:
\begin{equation}
\Delta_{\rm C} E_{\rm aver} = \frac {5\pi \alpha^3  }{32 m^2}
\psi^2(0)
\left( \frac {1}{\epsilon} - 4 \ln(m\alpha)+\frac {7}{5} \right).
\end{equation}

Virtual transitions to negative energy states induced by the Coulomb
exchanges, Fig.~\ref{fig:soft}(h), generate an effective
spin-independent operator
\begin{equation}
V_{C-}(\r) = - \frac{1}{4m^3} [\p,C(r)]^2.
\end{equation}
This operator describes the energy shift due to a creation
of an additional $e^+ e^-$ pair by the Coulomb field of either electron
or positron.  The resulting energy shift is
\begin{equation}
\Delta_{C-} E_{\rm aver} = - \frac {\pi \alpha^3  }{4m^2}
\psi^2(0)
\left (
\frac {1}{\epsilon} - 4 \ln(m\alpha) \right ).
\end{equation}

The spin-independent part of the tree level magnetic exchange,
Fig.~\ref{fig:soft}(b), induces the following shift in the energy
levels:
\begin{equation}
\Delta_{\rm M} E_{\rm aver} =  \frac{2\pi\alpha}{m^4}
\left\langle (\p'^2+\p^2)  
\left( {\p\p'\over \q^2}- {(\p\q)(\q\p')\over \q^4}\right)
\right\rangle 
= \frac {\pi \alpha^3}{2 m^2}\psi^2(0)
\left (  \frac {1}{\epsilon} - 4 \ln(m\alpha)+\frac {1}{2} \right ).
\end{equation}

To account for the retardation in the magnetic photon propagation,
Fig.~\ref{fig:soft}(c,d,e), we use the approach described in the HFS
case. Our starting point is similar to Eq.~(\ref{retar}), except that
now the full expression for the currents must be used, rather than
just their spin-dependent part. We obtain
\begin{equation}
\Delta_{\rm ret} E_{\rm aver} = \frac {\pi \alpha^3}{8m^2}
\psi^2(0)
\left (
\frac {1}{\epsilon} - 4 \ln(m\alpha) -8  \right ).
\end{equation}

The next contribution comes from the exchange of 
two magnetic photons with creation of an additional
$e^+ e^-$ pair in the intermediate state, Fig.~\ref{fig:soft}(i).
We find
\begin{equation}
\Delta_{\rm MM-} E_{\rm aver} = -\frac {\pi \alpha^3}{2 m^2}
\psi^2(0)
\left (
\frac {1}{\epsilon} - 4 \ln(m\alpha) -3 \right ).
\end{equation}

We proceed further with the correction to $E_{\rm aver}$,
induced by the second iteration of the $S$-wave 
Breit Hamiltonian. The calculation closely follows 
the HFS case. We arrive at the following result:
\begin{equation}
\Delta_{\rm S}^{(2)} E_{\rm aver} = - \frac {\pi \alpha^3}{12 m^2}
|\psi(0)|^2
\left (
\frac {1}{\epsilon} - 4 \ln(m\alpha) + \frac {433}{24} \right ).
\end{equation}

The iteration of the $D$-wave part of the Breit Hamiltonian only
influences the energy levels of the triplet state because of the total
angular momentum ($\vec{L}+\vec{S}$) conservation.  For this reason, to
obtain the required correction to $E_{\rm aver}$ it is sufficient to
multiply Eq.~(\ref{DeltaD2}) by the factor $d/(d+1)$.  We find
\begin{equation}
\Delta_D^{(2)} E_{\rm aver} = \frac {5 \pi \alpha^3}{96 m^2}
\psi^2(0)
\left (
\frac {1}{\epsilon} - 4 \ln(m\alpha) - \frac {119}{30} \right ).
\end{equation}

It is easy to see that in 3 dimensions the spin-dependent operators do
not contribute to $E_{\rm aver}$.  However, since we work with 
divergent integrals and use dimensional regularization, this is no
longer valid for $d \ne 3$.  In this case an ``anomalous'' situation
arises: spin-dependent operators provide contributions of the form
$(d-3)/\epsilon$ to $E_{\rm aver}$, which are finite as $\epsilon \to 0$.  Part
of these contributions has already been accounted for in the corrections
induced by the Breit Hamiltonian.  The remaining contributions give
\begin{equation}
\Delta_{\rm anom} E_{\rm aver} = 
- \frac {15}{64} \frac {\pi \alpha^3 }{m^2}\psi^2(0).
\end{equation}

The hard scale contribution, Fig.~\ref{fig1}, is calculated in the
same way as for the HFS. One finds:
\begin{equation}
\Delta_{\rm hard} E_{\rm aver} = -\frac {\pi \alpha^3 }{3m^2}
        \psi^2(0)
       \left (  \frac {13}{8} + \frac {9 \zeta (3)}{\pi^2}
         +\frac {33}{2\pi ^2} \right ).
\end{equation}

The sum of all contributions presented above provides the 
${\cal O}(m\alpha^6)$ pure recoil correction to the ground state energy:
\begin{equation}\label{n=1shift}
\Delta_{\rm rec} E_{\rm aver} = -\frac {m\alpha^6}{8} 
\left ( \frac {901}{576}  +\frac {11}{2\pi^2}
      + \frac {3 \zeta (3)}{\pi ^2}    \right ) = -\frac
{m\alpha^6}{8} (2.48688\ldots),
\end{equation}
in very good agreement with the numerical result of Eq.~(20) in
\cite{PhPRL}, $-\frac {m\alpha^6}{8} (2.484(5))$. 

\subsection{Energy levels for arbitrary $n$}

To generalize the result Eq.~(\ref{n=1shift}) for arbitrary $n$, we
proceed according to the program outlined in Section \ref{sec:nhfs}.  We
repeat the calculation of the soft-scale contributions to $E_{\rm
aver}$ for arbitrary $n$ using a different regularization scheme.
Namely, we set $d=3$ and cut off the divergent integrals over $r$ from
below at some $r_0 \ll 1/(m\alpha)$.  The transition to three
dimensions simplifies the calculation.  We find
\begin{equation}
\Delta_{\rm rec} E_{\rm aver} (n) = 
-\frac {m \alpha^6}{8n^3} \left ( [div] +\frac {69}{64n^3}
-\frac {8}{3n^2}+\frac {2}{n} \right ).
\label{Eprel}
\end{equation}
The $n$-independent term in the above equation is regularization
dependent. It cannot be determined by considering the soft-scale
contributions alone.  Nevertheless, by matching Eq.~(\ref{Eprel}) to
the shift for the ground state Eq.~(\ref{n=1shift}), the ``value'' of
the divergent constant $[div]$ is completely determined. We obtain
\begin{equation}
\Delta_{\rm rec} E_{\rm aver} (n) = -\frac {m \alpha^6}{8n^3} \left (
\frac {83}{72} + \frac {11}{2\pi^2} + \frac {3 \zeta (3)}{\pi ^2}
+\frac {69}{64n^3} -\frac {8}{3n^2}+\frac {2}{n}
\right ).
\label{Eaverfin}
\end{equation}
This is our main result for the recoil corrections to the energy
levels of positronium. It agrees with the partially numerical result
derived in \cite{PhPRL}.

\section{Radiative recoil corrections}

\begin{figure} 
\hspace*{-2mm}
\begin{minipage}{16.cm}
\vspace*{3mm}
\[
\hspace*{-5mm}
\mbox{ 
\begin{tabular}{cc}
\psfig{figure=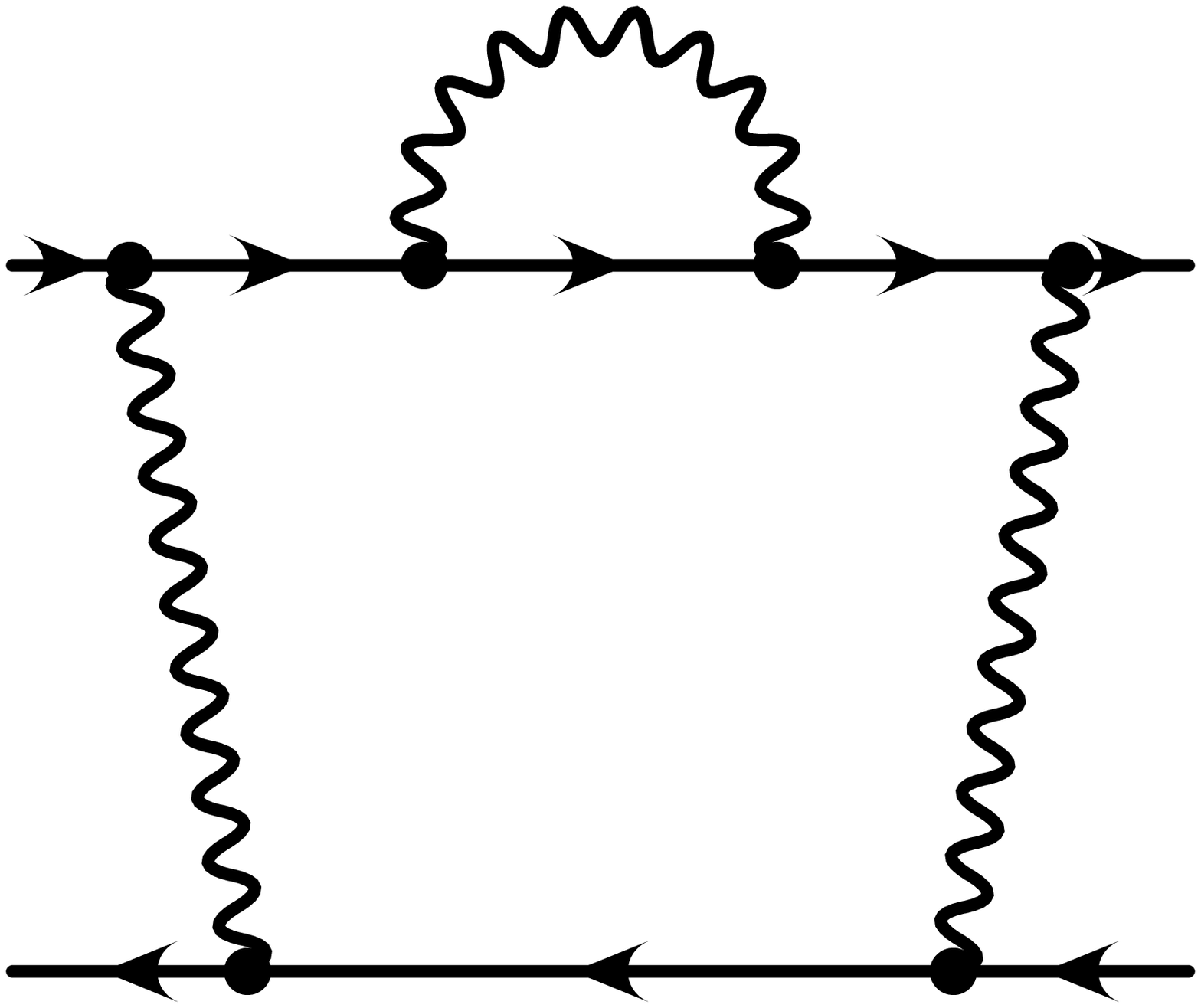,width=30mm,bbllx=72pt,bblly=291pt,%
bburx=544pt,bbury=530pt} 
& \hspace*{10mm}
\psfig{figure=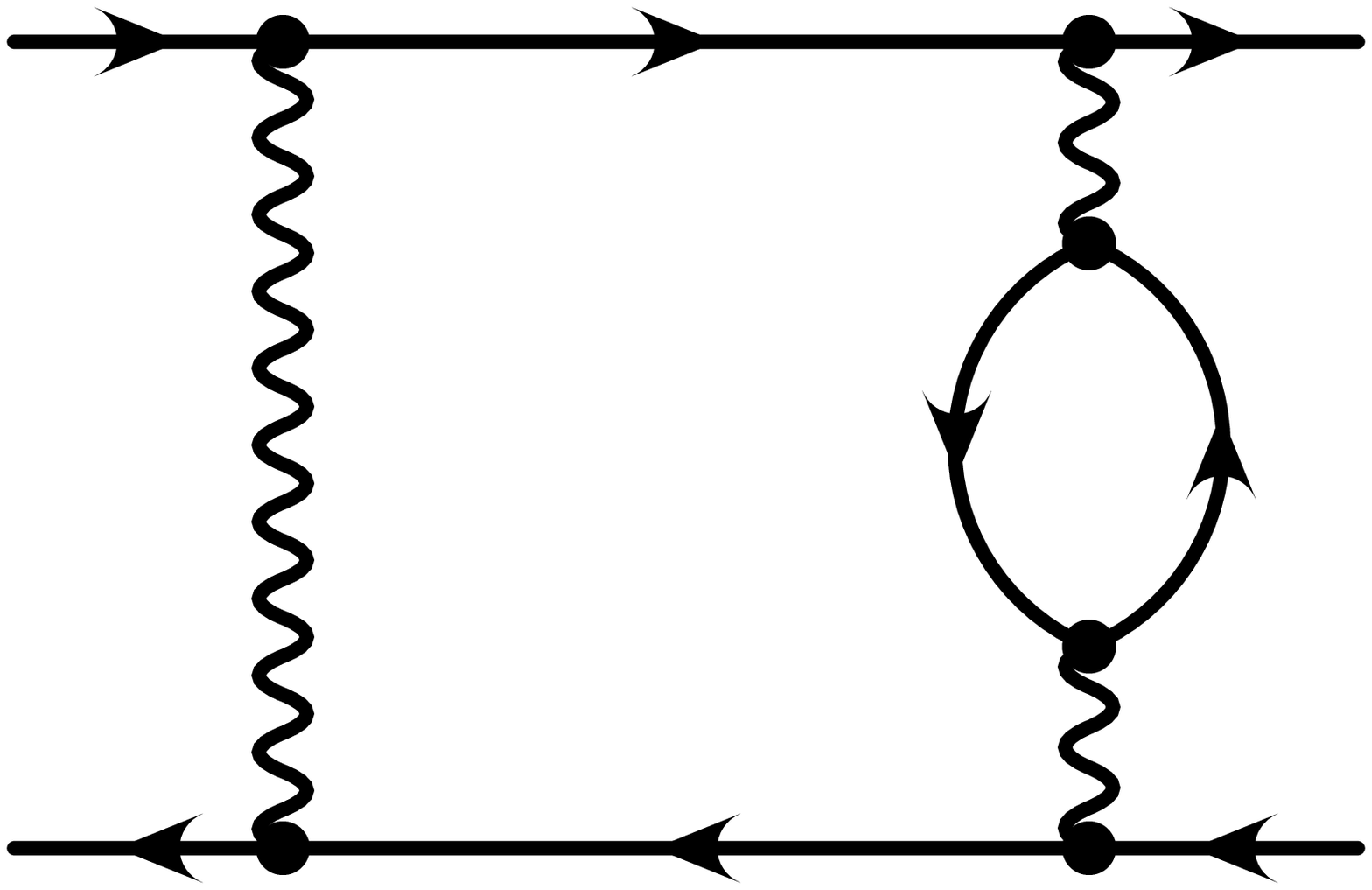,width=30mm,bbllx=72pt,bblly=291pt,%
bburx=544pt,bbury=530pt} 
\\[12mm]
\psfig{figure=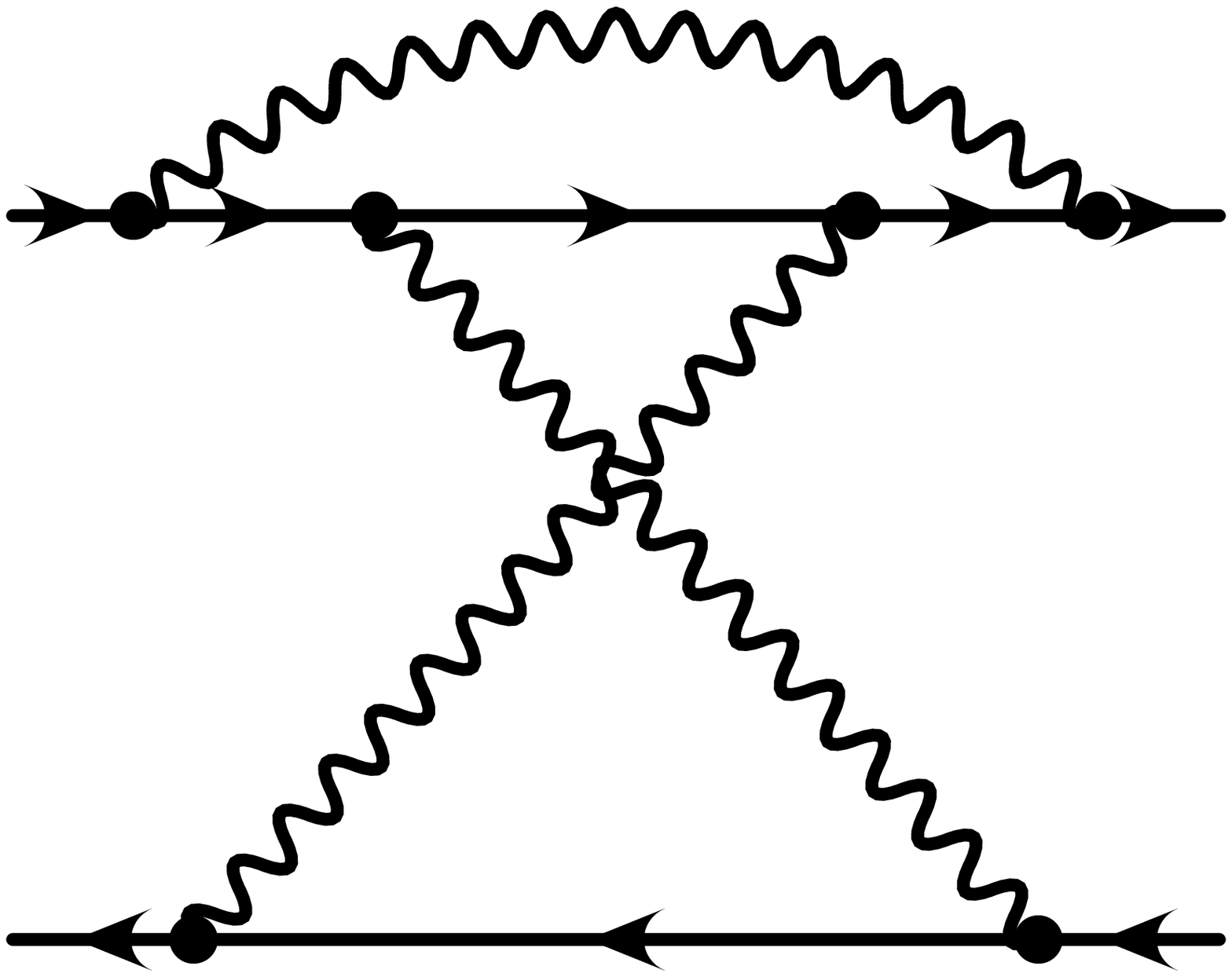,width=30mm,bbllx=72pt,bblly=291pt,%
bburx=544pt,bbury=530pt} 
& \hspace*{10mm}
\psfig{figure=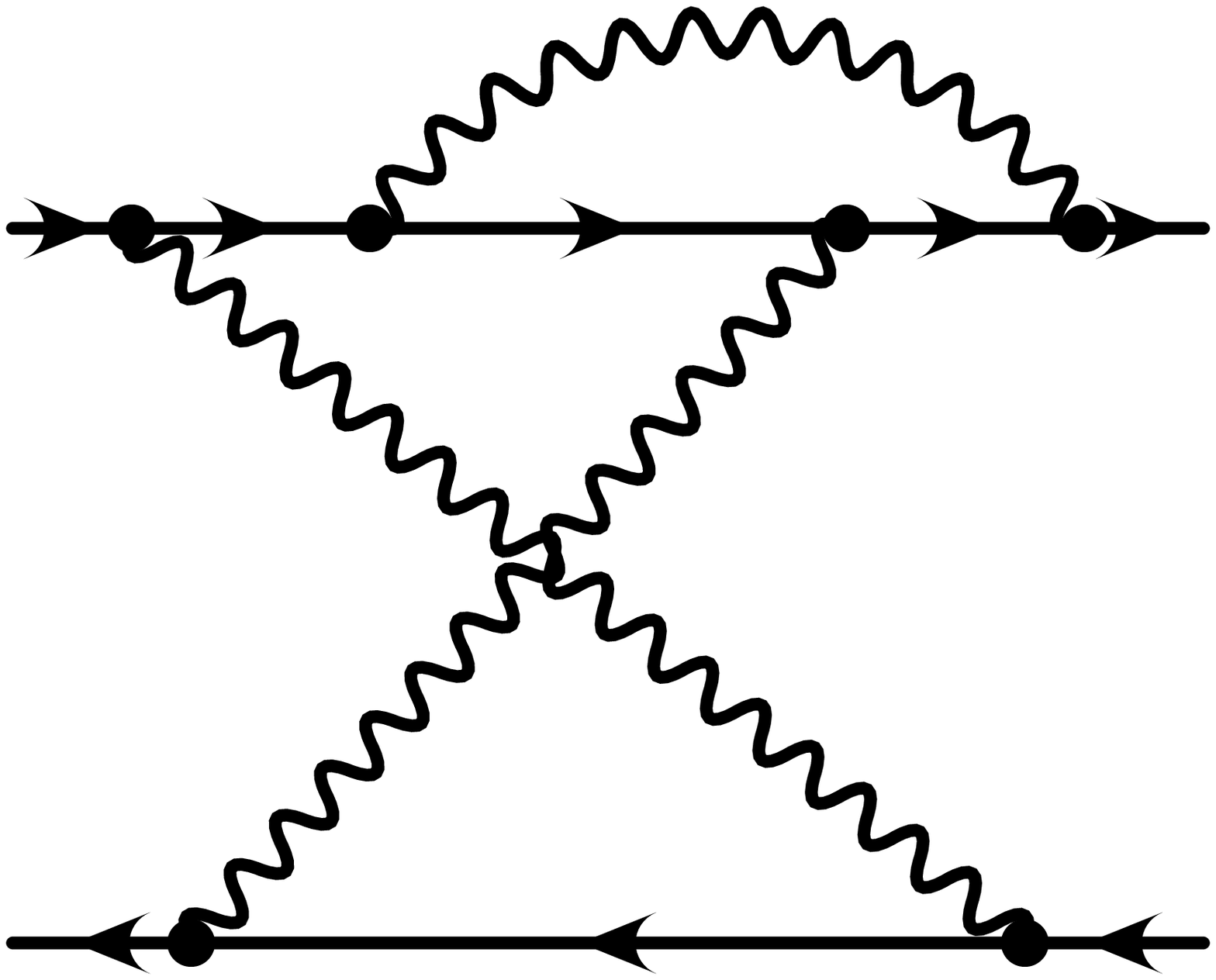,width=30mm,bbllx=72pt,bblly=291pt,%
bburx=544pt,bbury=530pt} 
\end{tabular}
}
\]\vspace*{6mm}
\end{minipage}
\caption{Examples of radiative recoil corrections to
positronium HFS and spin-averaged energy levels.}
\label{fig2}
\end{figure}

So far in this paper we have been considering pure recoil effects.
Another class of the ${\cal O}(m\alpha^6)$ corrections
to positronium energy levels and their HFS are the so-called
radiative recoil corrections, where one of the  three exchanged 
photons is created and absorbed by the same particle
(see Fig.~\ref{fig2}).

Our technique is very convenient for the calculation of these
corrections. The key point is that at ${\cal O}(m \alpha^6)$  the
radiative recoil corrections do not receive any contribution from
the non-relativistic scales. Thus it is sufficient to calculate the
diagrams shown in Fig.~\ref{fig2} (supplemented by the electric
charge, electron wave function and mass renormalization) exactly at
the threshold.  For the same reason, the
$n$-dependence of the radiative recoil corrections comes only from the
$1/n^3$ behavior of the $nS$-wave function at the origin. 
Some details of this calculation are described in Section
\ref{sec:hard} in the context of the HFS.  We obtain 
\begin{eqnarray}
\Delta_{\rm rad\; rec} E_{\rm hfs} 
&=& \frac {m\alpha^6}{n^3} \left [ \frac {\zeta(3)}{2\pi^2}
-\frac{79}{48}+\frac{41}{36\pi^2} +\frac {4}{3}\ln 2 \right ],
\label{radrechfs}
\\
\Delta_{\rm rad\; rec} E_{\rm aver} &=& \frac {m\alpha^6}{n^3}
\left[
\frac {9\zeta (3)}{8\pi^2} + \frac {97}{144}-\frac {1025}{432\pi^2}
\right],
\label{radreclevels}
\end{eqnarray}
respectively for corrections to the HFS and to the average energy, in
full agreement with the analytic results of Ref.~\cite{PhK}.
For completeness, we give here separately the contributions of 
electron vacuum polarization effects to radiative recoil corrections
\cite{Eides95,PhK,Sapirstein:1984xr}
(they are included in (\ref{radrechfs},\ref{radreclevels})):
\begin{eqnarray}
\Delta_{\rm rad\; rec}^{\rm vac\;pol} E_{\rm hfs} 
=\frac {m\alpha^6}{n^3} {5\over 9\pi^2},
\qquad
\Delta_{\rm rad\; rec}^{\rm vac\;pol} E_{\rm aver} = \frac {m\alpha^6}{n^3}
\left( {1\over 36}-{5\over 27\pi^2}\right).
\end{eqnarray}

\section{Summary and Conclusions}
\label{sec:sum}
The main new results of the present paper are the analytic formulas
(\ref{Ehfsrecn}) and (\ref{Eaverfin}) for the pure recoil ${\cal
O}(m\alpha^6)$ corrections to the HFS and spin--averaged energy levels
of positronium $nS$ states.  These recoil effects provide the
last pieces needed to present complete analytical formulas for the
total corrections to $E_{\rm aver}$ and $E_{\rm hfs}$.
We use the parameterization introduced in
Eq.~(\ref{param}),
\begin{equation}
E(J,n) = E_{\rm aver}(n) +\left( {1\over 4} - \delta_{J0} \right)
E_{\rm hfs}(n),
\end{equation}
and find
\begin{eqnarray}
E_{\rm aver}(n)&=& -\frac {m\alpha^2}{4n^2}
+\frac {m\alpha^4}{16n^3} \left (\frac {11}{4n} - 1 \right )  
\nonumber \\
&&+\frac {m\alpha^5}{8\pi n^3} 
\left[ 
-6\ln \alpha 
-\frac {16}{3} \ln k_0(n,0)
+ \frac {14}{3} \left( \ln \frac {4}{n} + \Psi(n) + \gamma_E \right) 
-\frac {37}{45} - 3 \ln 2 + \frac{7}{3n} 
\right ]
\nonumber \\
&&+\frac{m\alpha^6}{32n^3} \left[
-\ln\frac{\alpha}{n} - \Psi(n) - \gamma_E
+ \frac{141}{4}\frac{\zeta(3)}{\pi^2} 
+ \left( \frac{137}{6} - \frac{68}{\pi^2} \right) \ln 2
+ \frac{1421}{27\pi^2} \right. \nonumber \\
&&\qquad \qquad \left. - \frac{2435}{432} - \frac{7}{n}
+ \frac{17}{12n^2} - \frac{69}{16n^3} \right],
\label{aver:final}
\end{eqnarray}
and
\begin{eqnarray}
E_{\rm hfs}(n)&=&
{7\over 12}\frac {m\alpha^4}{n^3} 
-\frac {m\alpha^5}{\pi n^3} \left( \frac {8}{9} + \frac {1}{2} \ln 2 \right) 
+\frac {m\alpha^6}{n^3} 
\left [
-\frac {5}{24} \left ( \ln \frac {\alpha}{n} + \Psi(n) + \gamma_E \right )
+\frac {1367}{648\pi^2}
\right. 
\nonumber \\  && 
\left.
 -\frac {4297}{3456} +
\left (\frac {221}{144} + \frac {1}{2\pi^2 } \right ) \ln 2
-\frac {53}{32\pi^2 } \zeta (3)
+ \frac {5}{8n}-\frac {85}{96n^2}
\right].
\label{hfs:final}
\end{eqnarray}
We have also recalculated the radiative recoil corrections,
Eqs.~(\ref{radrechfs},\ref{radreclevels}), confirming recent result of
Ref.~\cite{PhK}.  Let us make a technical remark.  In dimensional
regularization, used in this paper, the calculation of the radiative
recoil corrections is particularly simple. Since there are no
low-scale contributions to the radiative recoil corrections, it
suffices to calculate corresponding Feynman graphs exactly at the
threshold.  No matching or subtractions are required.

Formulas (\ref{aver:final},\ref{hfs:final}), together with $P$ state
energy levels given in Appendix \ref{app:P}, can be used to compute
quantities which can be directly confronted with experimental data.
We use the following values for the Rydberg \cite{Hansch} and fine
structure \cite{Czarnecki:1998nd} constants:
\begin{equation}
R_\infty =\frac{m\alpha^2}{2}=3\,289\,841\,960.394(27)\;{\rm MHz},
\;\;\;\;\; \alpha=1/137.035\,999\,59(51).
\end{equation}
In addition to the full corrections ${\cal O}(m\alpha^6)$ we include
the leading logarithmic terms  ${\cal O}(m\alpha^7\ln^2\alpha)$ found
in \cite{DL} for HFS, and in \cite{KirSash} for the spin-averaged energy
levels:
\begin{equation}
\Delta_{LL} E(J,n) = - \left( \frac{499}{15}+7\left(1-4\delta_{J0}\right) \right)
\frac{m\alpha^7 \ln^2 \alpha}{32\pi n^3} \delta_{l0}.
\label{eq:llogs}
\end{equation}

For the most precisely measured quantity, the Ps ground state HFS, we
find 
\begin{equation}
\label{HFSth}
\Delta \nu =  203\,392.01(46)   \; {\rm MHz}.
\end{equation}
The ${\cal O}(m\alpha^6)$ recoil corrections to this observable have
been subject of some debate.  In the literature three different
results have been reported \cite{Caswell:1986ui,Ph,AS}.\footnote{After
our HFS calculation was completed, we were informed about an
independent numerical calculation of the recoil corrections \cite{AB}.
Although that study is still in progress, its preliminary results seem
to agree with Ref.~\cite{Ph} and the present paper.}  Our result for
this correction, Eq.~(\ref{Ehfsrecground}), evaluates numerically to $
m\alpha^6 \left ( - \frac{ 1 }{ 6 } \ln \alpha + 0.37632 \right )$.
This is in excellent agreement with Ref.~\cite{Ph}, where for the
non-logarithmic part of the correction a number 0.3767(17) was
obtained.  The framework of our calculation is similar to
Ref.~\cite{Ph}.  However, in that study a different regularization
method was used.  The agreement of the results gives us confidence in
their correctness.

Comparing Eq.~(\ref{HFSth}) with the experimental results,
Eqs.~(\ref{Mills},\ref{Hughes}), we observe a significant deviation of
the order of $3-4$ experimental errors.  It is not very likely that
the uncalculated higher order effects alone can account for this
discrepancy.  The size of the ${\cal O}(m\alpha^6)$ corrections gives
no indication of bad behavior of the perturbative expansion.  On the
other hand, the leading logarithmic term ${\cal O}
(m\alpha^7\ln^2\alpha)$ is sizable.  A calculation of the subleading
terms at this order remains an important theoretical challenge.

For another experimentally interesting quantity, the energy interval
of the $1S-2S$ transition, we get
\begin{equation}\label{Edifth}
E(2^3S_1)-E(1^3S_1) =1233\,607\,222.18(58) \; {\rm MHz}.
\end{equation}
in fair agreement with the experimental result, Eq.~(\ref{Fee}).

Other quantities, for which high precision measurements have been made
or are being planned, have recently been reviewed in \cite{PhK}.  In
Table \ref{tab:one} we update the theory predictions for those
observables.  Our predictions are in good agreement with \cite{PhK}.
We have been able to decrease the error bars by including the
analytical results (\ref{aver:final},\ref{hfs:final}) and the value of the
leading quadratic logarithms (\ref{eq:llogs}). 

Finally we would like to comment on our error estimates.  The errors
due to uncertainties in the fine structure constant and the electron
mass are well below $0.1$ MHz level.  The dominant theoretical error
source is the uncalculated remainder of the perturbation expansion.
Although formally $m\alpha^7 \sim 0.1\; {\rm MHz}$, the leading ${\cal
O}(m\alpha^7 \ln^2 \alpha)$ terms contribute $-0.92\; {\rm MHz}$ to
the HFS \cite{DL}.  It remains very important to calculate the
remaining, non-leading terms in ${\cal O} (m\alpha^7)$.
For the present analysis we assume that the leading logs ${\cal
O}(m\alpha^7 \ln^2 \alpha)$ dominate the higher order contributions
and take half their size as the theoretical error estimate.

The spectrum of the $nS$ and $nP$ positronium energy levels is now
known analytically, including effects ${\cal O}(m\alpha^6)$.  Our
calculation for the $nS$ levels was made possible by new theoretical
tools which have their roots in the recent perturbative calculations
in high-energy physics.  We hope that these methods will find further
applications.

The agreement between theoretical predictions and experimental results
in Ps spectroscopy is impressive with a few exceptions.  One can only
hope to find something new and unexpected by trying to put these
exceptions in line with the overall picture.  We look forward to
future improved measurements of positronium energy levels and their
confrontation with QED.

\subsection*{Acknowledgments}

We are grateful to A.~Burichenko for informing us about his results
prior to publication.  We thank S. Karshenboim, K. Pachucki and
E. Remiddi for reading the manuscript and helpful comments.
K.M.~would like to thank the High Energy Theory Group at Brookhaven
National Laboratory for hospitality during the final stage of this
project.  This research was supported in part by the U.S.~Department
of Energy under grant number DE-AC02-98CH10886, by BMBF under grant
number BMBF-057KA92P, by Gra\-duier\-ten\-kolleg
``Teil\-chen\-phy\-sik'' at the University of Karlsruhe and by the
Russian Foundation for Basic Research under grant number 99-02-17135.

\appendix
\section{Tree--level electron--positron potential}
We present here formulas for the potential arising from a
single Coulomb or magnetic photon exchange between an electron and a
positron, valid to ${\cal O}(v^4)$.  
The virtual
annihilation is not taken into account here. We 
also drop those terms which annihilate the $S$--state wave function.
These formulas, valid in $d$-dimensions,  are
useful in the derivations of HFS and spin--averaged energy levels.  

For a single Coulomb exchange between two particles of opposite
charges, Fig.~\ref{fig:soft}(a), the minus on--shell scattering
amplitude is
\begin{equation}\label{AC}
- A_C(\p',\p) = - \frac{4\pi\alpha}{\q^2} \rho (\pp, \p) \rho (-\pp, -\p),
\end{equation}
where $\p$ and $\pp$ are spatial momenta of  the incoming and outgoing
electron; $\q= \pp - \p$; and the charge density is
$
\rho (\pp, \p) = u^+(\pp) u(\p)
$.
In momentum representation a single Coulomb photon exchange gives rise
to the potential
\begin{eqnarray}
U_C(\p,\p') &=& -{4\pi\alpha\over \q^2}\left\{
1-{\q^2\over 4m^2} 
+{5\left(\p^2-\p'^2\right)^2 + 6\q^2\left(\p^2+\p'^2\right)+\q^4
 +[\si\p',\si\p][\si'\p',\si'\p] \over 64m^4}\right\}.
\nonumber \\
\label{eqa:coul}
\end{eqnarray}
In the leading nonrelativistic approximation
Eq.~(\ref{eqa:coul}) gives the Coulomb potential.

Next we consider a magnetic photon exchange, Fig.~\ref{fig:soft}(b).
We neglect retardation effects.  The scattering amplitude is
\begin{equation}
- A_M(\p',\p) =  \frac{4\pi\alpha}{\q^2} j_i (\pp, \p) j_j (-\pp, -\p) 
                 \left( \delta_{ij} - \frac{q_i q_j}{\q^2} \right),
\end{equation}
where
$\vec{j}(\pp, \p) = u^+(\pp)\vec{\alpha} u(\p) $
is the matrix element of the current.
The resulting potential is
\begin{eqnarray}
U_M(\p,\p') &=&{\pi\alpha\over m^2 \q^2 }
 \left( {4\over \q^2}\left[ (\p\p')^2 - \p^2\p'^2\right]
 -{1\over 4}[\si\q,\sigma^i][\si'\q,\sigma'^i]\right)
\nonumber \\
&& 
-{\pi\alpha\over 2m^4\q^2}\left\{
(\p^2+\p'^2)
 \left( {4\over \q^2}\left[ (\p\p')^2 - \p^2\p'^2\right]
 -{1\over 4}[\si\q,\sigma^i][\si'\q,\sigma'^i]\right)
\right.
\nonumber \\
&& 
\left. 
\qquad\qquad
+ {\p^2-\p'^2\over 16}\left( [\si\q,\sigma^i][\si'\PP,\sigma'^i]
+[\si\PP,\sigma^i][\si'\q,\sigma'^i]\right)
\right\}.
\label{eqa:magn}
\end{eqnarray}
These formulas are valid in the center of mass frame.
We use $\p$ and $\p'$ to denote incoming and outgoing electron
momenta, and $\q\equiv \p'-\p$, $\PP\equiv \p'+\p$.  The primed
$\sigma$-matrices act on the positron spinor.

\section{Useful integrals}
\label{app:int}
In this Appendix we present various integrals which were useful in the
calculations presented in this paper.  The following formulas have
been used throughout the paper, especially for the tree level
diagrams: 
\begin{eqnarray}
\left\langle \p^2 \right\rangle &=& -{ m^2\alpha^2 \over 4}\psi^2(0),
\nonumber \\
\left\langle {\p^4 \over \q^2 } \right\rangle &=& \phantom{-} { m^2\alpha^2 \over 16 }\psi^2(0),
\nonumber \\
\left\langle {\p^2 {\p'}^2 \over \q^2 } \right\rangle &=&
 \phantom{-} { m^2\alpha^2 \over 4 }\psi^2(0)
\left( {1\over \epsilon } -4\ln (m\alpha) +{1\over 4}\right),
\nonumber \\
\left\langle {(\q\p)(\q\p')\over \q^2 } \right\rangle &=& -{ m^2\alpha^2 \over 8 }
\psi^2(0)\left( {1\over \epsilon } -4\ln (m\alpha) - 1\right).
\end{eqnarray}

In the remainder of this Appendix we describe some details of the
$D$-wave contribution to the second iteration of the Breit
Hamiltonian.  
First, we rewrite Eq.~(\ref{D02}) in the following way:
\begin{eqnarray}
\lefteqn{\left\langle B_{ij}(\p'-\k)g(k)B_{ij}(\k-\p) \right\rangle
=
\left\langle B_{ij}(\p'-\k)g(k)4\pi\frac{k_i k_j - 2k_i p_j + p_i p_j
}{ (\k-\p)^2 } \right\rangle}
 \nonumber \\
&&\qquad=
\left\langle B_{ij}(\q)p_i p_j
- 2 B_{ij}(\p'-\k)g(k)\frac{4\pi(k_i-p_i)}{(\k-\p)^2} p_j 
- B_{ij}(\p'-\k)g(k)\frac{4\pi}{(\k-\p)^2} p_i p_j \right\rangle.\label{0mod}
\end{eqnarray}
Here and below we use the Schr\"odinger equation in the form
\begin{equation}
\phi(p) = g(p)\frac{4\pi}{(\p-\k)^2} \phi(k).
\end{equation}
Similarly, by rearranging terms in (\ref{D12}) we get
\begin{eqnarray}
\lefteqn{\left\langle B_{ij}(\p'-\k')g(k')
   \frac{4\pi}{(\k'-\k)^2}g(k)B_{ij}(\k-\p) \right\rangle}
   \nonumber \\
   &=&
   \left\langle B_{ij}(\p'-\k)g(k)\frac{4\pi}{(\k-\p)^2} p_i p_j \right.
   -2 B_{ij}(\p'-\k')g(k')
   \frac{4\pi}{(\k'-\k)^2}k_i g(k)\frac{4\pi}{(\k-\p)^2}p_j
   \nonumber \\
   && \left. \qquad
   +B_{ij}(\p'-\k')g(k')
   \frac{4\pi}{(\k'-\k)^2}g(k)\frac{4\pi}{(\k-\p)^2}p_i p_j \right\rangle,
\end{eqnarray}
Using the symmetry with respect to $\p\leftrightarrow\p'$, we rewrite
the first term in Eq.~(\ref{0mod}),
\begin{equation}\label{lin}
\left\langle B_{ij}(\q)p_i p_j \right\rangle = 4\pi \left\langle \frac{\q^2}{2}
                  + \frac{(\p'\q)(\q\p)}{\q^2} - \frac{\p^2}{3} \right\rangle 
                  = 4\pi \left\langle - \frac{2}{3}\gamma^2
                  + \frac{(\p'\q)(\q\p)}{\q^2} \right\rangle.
\end{equation}
In the same way, the second term in Eq.~(\ref{0mod}) is transformed
to
\begin{eqnarray}
&&\left\langle - 2 B_{ij}(\p'-\k)g(k)\frac{4\pi(k_i-p_i)}{(\k-\p)^2} p_j \right\rangle
\nonumber \\
&&\qquad\qquad =\left\langle \frac{8\pi}{ d } g(k)\frac{4\pi(\k-\p)\p}{(\k-\p)^2} 
+ 2 p'_i\frac{4\pi}{(\p'-\k)^2}
\left( 2k_j - p'_j \right) g(k)\frac{4\pi(k_i-p_i)}{(\k-\p)^2} p_j \right\rangle.
\end{eqnarray}
Considering the divergent part of this expression we find
\begin{equation}
\left\langle \frac{8\pi}{ d } g(k)\frac{4\pi(\k-\p)\p}{(\k-\p)^2} \right\rangle
= \frac{16\pi^2\alpha}{ d }\psi^2(0) \left[ G_0(0,0) + G_1(0,0) \right].
\end{equation}
The sum of (\ref{D02}) and (\ref{D12}) reads 
\begin{eqnarray}\label{D2}
\Delta_{D0} E_{\rm hfs} + \Delta_{D1} E_{\rm hfs} 
&=& - \frac{ \alpha(d-2)^2 }{ 4m^4d }
                  \left\langle \frac{16\pi^2\alpha}{ d }
                  \left[ G_0(0,0) + G_1(0,0) \right]
                  - \frac{8\pi\gamma^2}{3}
                  + \frac{4\pi(\p'\q)(\q\p)}{\q^2} \right.
                  \nonumber \\
                  &&
                  + 2 p'_i\frac{4\pi}{(\p'-\k)^2}
\left( 2k_j - p'_j \right) g(k)\frac{4\pi(k_i-p_i)}{(\k-\p)^2} p_j
                  \nonumber \\
                  &&
   -2 B_{ij}(\p'-\k')g(k')
   \frac{4\pi}{(\k'-\k)^2}k_i g(k)\frac{4\pi}{(\k-\p)^2}p_j
   \nonumber \\
   && \left.
   +B_{ij}(\p'-\k')g(k')
   \frac{4\pi}{(\k'-\k)^2}g(k)\frac{4\pi}{(\k-\p)^2}p_i p_j \right\rangle.
\end{eqnarray}
Only two terms here contain the logarithmic divergence:
\begin{eqnarray}\label{Dlog1}
- \frac{ 4\pi^2\alpha^2(d-2)^2 }{ m^4d^2 }\left\langle G_1(0,0) \right\rangle
&=& \frac{ \pi\alpha^3 }{ 36 m^2 } \psi^2(0)
               \left( \frac{ 1 }{\epsilon } - 4 \ln(m\alpha) - \frac{2}{3} \right),
               \\
- \frac{ \alpha(d-2)^2 }{ 4m^4d }
                  \left\langle \frac{4\pi(\p'\q)(\q\p)}{\q^2} \right\rangle
&=& \frac{ \pi\alpha^3 }{ 24 m^2 } \psi^2(0)
               \left( \frac{ 1 }{\epsilon } - 4 \ln(m\alpha) - \frac{13}{3} \right).
\end{eqnarray}
All other terms are finite and we compute them in three dimensions.
Here we list some useful integrals ($x\equiv
k/\gamma$ and $a(x)\equiv\arctan(x)$): 
\begin{eqnarray}
&&\int {{\rm d}^3 p\over (2\pi)^3} \frac{p_i}{(\k-\p)^2} \phi(p) =
      \frac{ \gamma\psi(0) k_i}{k^3}
      \left(a(x) - \frac{x}{x^2+1}
      \right), \nonumber
      \\
&&\int {{\rm d}^3 p\over (2\pi)^3} \frac{p_i p_j}{(\k-\p)^2}\phi(p) 
\nonumber\\
&&\qquad \qquad 
=
      \frac{ \gamma\psi(0)}{ k } \left[
      {1\over 2}\left( \frac{k_ik_j}{\k^2} - \frac{\delta_{ij}}{3} \right)
       \left({x^2+3\over x(x^2+1)} + {x^2-3\over x^2}a(x) \right)
      + \frac{\delta_{ij}}{3}\left(a(x) - \frac{x}{x^2+1}
      \right) \right],
\nonumber
      \\
&&\int {{\rm d}^3 p\over (2\pi)^3}  {{\rm d}^3 k'\over (2\pi)^3} 
\frac{g(k')B_{ij}(\k'-\p) }{(\k-\k')^2} \phi(p)
=      \frac{ \gamma\psi(0)}{ k }
      \left( \frac{k_ik_j}{\k^2} - \frac{\delta_{ij}}{3} \right)
     \left[ \frac{a(x)}{2} \left( 1 + \frac{3}{x^2} \right) - \frac{3}{2x} \right].
\nonumber \\
\end{eqnarray}
Using these above formulas in Eq.~(\ref{D2}) we find the final result
for $\Delta_{D0} E_{\rm hfs} + \Delta_{D1} E_{\rm hfs}$.

\section{$P$ state energy levels}
\label{app:P}
In this Appendix we present formulas 
for the energy levels of $P$  states,  to order ${\cal O}(m\alpha^6)$.
Correcting some minor misprints in Ref.~\cite{KMYp,KMYp2} one finds
\begin{eqnarray}
E(n^3P_2)&=&-\frac{m\alpha^2}{4n^2} 
-\frac{m\alpha^4}{4n^3}
 \left( \frac{13}{30}-\frac{11}{16n} \right) 
-\frac{m\alpha^5}{8\pi n^3}
\left( \frac{4}{45} + \frac{16}{3}\ln k_0(n,1) \right) \nonumber \\
&&+\frac {m\alpha^6}{n^3} \left(
-\frac {69}{512n^3}+\frac {559}{4800n^2}-\frac {169}{4800n}
+\frac {20677}{432000}-\frac {3}{80} \ln 2
+\frac {9\zeta(3)}{160\pi^2} + \frac {13}{128\pi^2}
\right),
\nonumber \\
E(n^3P_1)&=&-\frac{m\alpha^2}{4n^2} 
-\frac{m\alpha^4}{4n^3}
\left( \frac{5}{6}-\frac{11}{16n} \right) 
-\frac{m\alpha^5}{8\pi n^3}
\left( \frac{5}{9} + \frac{16}{3}\ln k_0(n,1) \right) \nonumber \\
&&+
\frac {m\alpha^6}{n^3} \left(
-\frac {69}{512n^3}+\frac {77}{320n^2}-\frac {25}{192n}
+\frac {1}{48}\ln 2  -\frac {\zeta(3)}{32\pi^2} -
\frac {179}{3456\pi^2} +\frac {493}{17280} \right),
\nonumber \\
E(n^3P_0)&=&-\frac{m\alpha^2}{4n^2} 
-\frac{m\alpha^4}{4n^3}
\left(  \frac{4}{3}-\frac{11}{16n} \right) 
-\frac{m\alpha^5}{8\pi n^3}
\left(  \frac{25}{18} + \frac{16}{3}\ln k_0(n,1) \right) \nonumber \\
&&+
\frac {m\alpha^6}{n^3} \left(
-\frac {69}{512n^3}+\frac {119}{240 n^2}-\frac {1}{3n}-
\frac {923}{4320} +\frac {1}{8}\ln 2 
-\frac {3}{16\pi^2}\zeta(3)-\frac {203}{576 \pi^2}
\right),
\nonumber \\
E(n^0P_1)&=&-\frac{m\alpha^2}{4n^2} 
-\frac{m\alpha^4}{4n^3}
\left(  \frac{2}{3}-\frac{11}{16n} \right) 
-\frac{m\alpha^5}{8\pi n^3}
\left(  \frac{7}{18} + \frac{16}{3}\ln k_0(n,1)
 \right) \nonumber \\
&&+
\frac {m\alpha^6}{n^3} \left(
\frac {163}{4320}+\frac {23}{120n^2}-\frac {69}{512n^3}
-\frac {1}{12n} \right).
\end{eqnarray}
For the numerical evaluations we  use the following values of  Bethe
logarithms $\ln\left[ k_0(n,l)/R_\infty \right]$ \cite{Drake90}
\begin{eqnarray}
\ln\left[ k_0(1,0)/R_\infty \right] &=&\phantom{-} 2.984\;128\;555\;765\;498,
\nonumber \\
\ln\left[ k_0(2,0)/R_\infty \right] &=&\phantom{-} 2.811\;769\;893\;120\;563,
\nonumber \\
\ln\left[ k_0(2,1)/R_\infty \right] &=&-0.030\;016\;708\;630\;213.
\end{eqnarray}


\begin{table}
\caption{Theoretical predictions for experimentally relevant
positronium transitions.}
\label{tab:one}
\begin{tabular}{ld}
Transition & Theory [MHz] \\ \hline
$2^3S_1 - 1^3S_1$ &    1233\,607\,222.18(58)    \\
$1^3S_1-1^1S_0$   &          203\,392.01(46)      \\
$2^3S_1-2^3P_0$   &    18498.25(8)       \\
$2^3S_1-2^3P_1$   &    13012.41(8)       \\
$2^3S_1-2^3P_2$   &     8626.71(8)      \\
$2^3S_1-2^1P_1$   &    11185.37(8)         \\
$2^3S_1-2^1S_0$   &  25424.67(6)         \\
\end{tabular}
\end{table}

\end{document}